\documentclass[12pt,preprint]{aastex}
\usepackage{psfig,natbib}
\shortauthors{Bower et al.}
\shorttitle{Circular Polarization of Sgr~A*}
\begin{document}

\newcommand\degd{\ifmmode^{\circ}\!\!\!.\,\else$^{\circ}\!\!\!.\,$\fi}
\newcommand{\etal}{{\it et al.\ }}
\newcommand{\uv}{(u,v)}
\newcommand{\rdm}{{\rm\ rad\ m^{-2}}}
\newcommand{\mylesssim}{\stackrel{\scriptstyle <}{\scriptstyle \sim}}

\slugcomment{Accepted for publication in the Astrophysical Journal}

\title{The Spectrum and Variability of Circular Polarization in Sagittarius
A* from 1.4 to 15 GHz}

\author{Geoffrey C. Bower\altaffilmark{1,2}, 
Heino Falcke\altaffilmark{3}, 
Robert J. Sault\altaffilmark{4} \&
Donald C. Backer\altaffilmark{2} }

\altaffiltext{1}{National Radio Astronomy Observatory, P.O. Box O, 1003 
Lopezville, Socorro, NM 87801; gbower@nrao.edu} 
\altaffiltext{2}{Astronomy Department \& Radio Astronomy Laboratory, 
University of California, Berkeley, CA 94720; gbower,dbacker@astro.berkeley.edu}
\altaffiltext{3}{Max Planck Institut f\"{u}r Radioastronomie, Auf dem 
H\"{u}gel 69, D 53121 Bonn Germany; hfalcke@mpifr-bonn.mpg.de} 
\altaffiltext{4}{Australia Telescope National Facility, P.O. Box 76,
Epping, NSW, 1710, Australia ; rsault@atnf.csiro.au}

\begin{abstract}

We report here multi-epoch, multi-frequency observations of the
circular polarization in Sagittarius A*, the compact 
radio source in the Galactic Center.  Data taken from the VLA
archive indicate that the fractional circular polarization at 4.8 GHz
was $-0.31\%$ with an rms scatter of $0.13\%$ from 1981 to 1998, 
in spite of a factor of 2
change in the total intensity.  The sign remained negative over the
entire time range, indicating a stable magnetic field polarity.
In the Summer of 1999 we obtained
13 epochs of VLA A-array observations at 1.4, 4.8, 8.4 and 15 GHz.
These observations employ a new technique that produces
an error of 0.05\% at 1.4, 4.8 and 8.4 GHz
and 0.1\% at 14.9 GHz.
In May, September and October of 1999 we obtained 11 epochs of 
Australia Telescope Compact Array observations at 4.8 and 8.5 GHz.
In all three of the data sets, we find no evidence for linear polarization
greater than 0.1\% in spite of strong circular polarization 
detections.
Both VLA and ATCA data sets support three conclusions regarding the fractional
circular polarization:  the average spectrum is
inverted with a spectral index $\alpha \approx 0.5 \pm 0.2$; 
the degree of variability is roughly constant on timescales of days to years; 
and, the degree of
variability increases with frequency. We also
observed that the largest increase in fractional circular polarization
was coincident with the brightest flare in total intensity.
Significant variability in the total intensity and fractional circular 
polarization on a timescale of 1 hour was observed during this flare,
indicating an upper limit to the intrinsic size during outburst of 70 AU at 15 GHz.
The fractional circular polarization at 15 GHz reached -1.1\%
and the spectral index is strongly inverted ($\alpha\sim 1.5$)
during this flare.  We tentatively conclude
that the spectrum has two components that match the high and low
frequency total intensity components.

\end{abstract}

\keywords{Galaxy: center --- galaxies: active --- polarization --- radiation
mechanisms: non-thermal --- scattering }

\section{Introduction}

The compact radio source in the Galactic Center, Sagittarius~A*,
is the best and closest candidate for a supermassive black hole in the
center of a galaxy 
\citep{1998ApJ...494L.181M,2001ARA&A..39..309M}.
The source Sgr A* is
positionally coincident with a $\sim 2.6 \times 10^6 M_{\sun}$ dark
mass 
\citep{2000MNRAS.317..348G,2000Natur.407..349G}.
Very long baseline interferometry (VLBI) has shown that
this source has a scale less than 1 AU and a brightness temperature in
excess of $10^9 {\rm\ K}$ 
\citep{1994ApJ...434L..59R,1998ApJ...496L..97B,1998ApJ...508L..61L,1998A&A...335L.106K,2001AJ....121.2610D}.
However, strong interstellar scattering of the
radiation along the line of sight has been shown to broaden the image
of Sgr~A* at radio through millimeter wavelengths 
\citep[e.g.,][]{1998ApJ...508L..61L,1994ApJ...427L..43F}.
As a consequence, VLBI observations have not convincingly demonstrated
the existence of source structure that would be an important
diagnostic of physical processes.
Long-term studies of Sgr~A*
indicate that the source shows no motion with respect to the center of
the Galaxy 
\citep{1999ApJ...524..805B,1999ApJ...524..816R}.
For these reasons, it is inferred that Sgr~A* is a
synchrotron or cyclo-synchrotron
emission region powered 
through accretion onto a supermassive black hole
\citep{1994ApJ...426..577M}. 
Significant details of the emission mechanism are not
understood.  In particular, leading models propose that the
emission originates in an advection dominated accretion flow, 
or ADAF \citep{2000ApJ...541..234O};
in a convection dominated accretion flow, or CDAF
\citep{2000ApJ...539..809Q};
or, in an outflow or jet \citep{1993A&A...278L...1F,2000A&A...362..113F}.

Polarization has proved to be an important tool in the study of AGN.
Studies of linear polarization (LP), which is typically on the order of a
few percent or less of the total intensity, have confirmed that the
emission process is synchrotron radiation and demonstrated that shocks
align magnetic fields in a collimated jet, leading to correlated 
variability in the total and polarized intensity 
\citep{1985ApJ...298..301H,1985ApJ...298..114M}.
Circular polarization (CP), on the other hand, is
less well understood in AGN.  Typically, the degree of CP
is $m_c < 0.1\%$ with only a few cases where $m_c$
approaches $0.5\%$ 
\citep{1983ApJS...52..293W,2000MNRAS.319..484R}.
The degree of CP usually
peaks near 1.4 GHz  and decreases strongly with increasing frequency.

Recently, VLBI imaging of several AGN has found 
localized CP
\citep{1999AJ....118.1942H}.
In 3C 279, these authors found
$m_{\rm c}\simeq1\%$ 
in an individual radio component with a fractional LP
of 10\% 
\citep{1998Natur.395..457W}.
The integrated CP, however, is less than 0.5\%. 
CP was also discovered in the radio jet of the
X-ray binary SS 433 at a level comparable to that of the LP
\citep{2000ApJ...530L..29F}.
The galactic microquasar GRS 1915+105 also shows strong
and variable CP that is associated with jet ejections 
\citep{2001MNRASprepfender}.
The CP in these sources
is probably produced through the conversion
of LP to CP by low-energy electrons in the 
synchrotron source.  This process is also known as repolarization 
\citep{1977OISNP..89.....P}.
CP has been identified through surveys in a number of high luminosity
AGN \citep{2000MNRAS.319..484R,2001ApJ...556..113H}.
A survey of low luminosity AGN found CP in only 1 of 11 galaxies,
M81* \citep{2001ApJ...560L.123B}.  This discovery is of particular
interest to this paper because of the similarity of M81* and 
Sgr A*, including the absence of LP.

In recent interferometric work we have
shown that the LP of Sgr~A* from centimeter to
millimeter wavelengths is extremely low. 
LP was not detected in a spectro-polarimetric
experiment with an upper limit of 0.2\% for
rotation measures as large as $10^7 \rdm$ at 8.4 GHz 
\citep{1999ApJ...521..582B}.
More recently, we have found that LP is less than 0.2\%
at 22 GHz and less than $\sim1\%$ at 112 GHz 
\citep{1999ApJ...527..851B,2001ApJ...555L.103B}.
Interstellar depolarization is very unlikely within the 
parameter space covered by these observations.  
\citet{2000ApJ...534L.173A} have claimed LP of
$\sim 10\%$ at $\nu \ge 150$ GHz on the basis of low resolution
JCMT observations.  If these results hold, then they have significant
implications for models of the emission mechanism 
\citep{2000ApJ...545..842Q,2000ApJ...538..L121,2000ApJ...545L.117M}.

Given these stringent limits on LP, the
presence of CP is not expected.  Nevertheless, we
have detected CP at a surprisingly
high level 
\citep{1999ApJ...523L..29B,1999ApJ...526L..85S}.
We found the fractional CP $m_c = -0.37 \pm 0.04 \%$
at 4.8 GHz and a spectral index $\alpha \sim -0.5$
between 4.8 and 8.4 GHz ($m_c \propto \nu^\alpha$).
Simple synchrotron models cannot produce the full polarization
characteristics without depolarization or repolarization in the
source or in the accretion region.  The introduction of low energy
electrons may lead to these Faraday effects and/or strong
gyrosynchrotron emission, which exhibits strong CP
\citep{1969ApJ...158..753R}.  Interstellar scattering in a
magnetized plasma may also lead to CP in radio sources
\citep{2000ApJ...545..798M}.

We describe here VLA and ATCA observations of CP and LP
in Sgr A*.  Because of the technical difficulties and the novelty
of these observations, especially for the VLA, we emphasize the
technique and the sources of error in our discussion.  In \S 2,
we discuss the analysis of 20 y of archival VLA data at 4.8 and
8.4 GHz.  These data show that the CP is stable
over this period.  In \S 3, we summarize new VLA observations in the
Summer of 1999 at 1.4, 4.8, 8.4 and 15 GHz.  We discuss sources
of error in these data in \S 4.  In \S 5, we introduce new ATCA
observations from 1999 at 4.8 and 8.5 GHz.  These observations
have very different systematic errors from those of the VLA.
We consider the long-term
evolution of CP in Sgr A* in \S 6 based on
results from the archival VLA data.  We analyse the new VLA and
ATCA data in \S 7 and \S 8.  This includes demonstration of consistency
between the various data sets and statistical measures of
variability.  We discuss our results in \S 9 and summarize
our conclusions in \S 10.

\section{VLA Archive Observations and Results}

We reduced data of Sgr~A* taken from the VLA archives.  All observations
were taken from the proper motion data sets of 
\citet{1999ApJ...524..805B}.
These observations had similar characteristics to those that we
reported in 
\citet{1999ApJ...523L..29B},
which were the final
epoch of the 
\citet{1999ApJ...523L..29B}
proper motion study.  Further
details of
the observations are given in 
\citet{1999ApJ...523L..29B}.

All observations were made in the A array.  
Eight epochs of observation were made, each consisting of two or
three five-hour runs spaced over a few weeks.
Data sets before 1989 include only 4.8 GHz observations. 
Data sets from 1989 onward include a small amount of 8.4 GHz observations.
Each data set has observations of Sgr~A*, 
W56 (J1745-2820), W109 (J1748-2907), GC 441 (J1740-2930),
J1743-0350 and J1751-2523.  Sgr~A*, W56, W109 and GC 441 were observed
multiple times within an hour.  J1743-0350 and J1751-2523 were
observed hourly.

Data reduction was performed with AIPS.  {\it A priori} amplitude
calibration was done using a single observation of 3C 286.  
Amplitude gains 
were determined with J1743-0350.  At 4.8 GHz, we also calibrated
the LP.  Polarization leakage terms were determined using
J1743-0350.  Polarization position angles were calibrated
with 3C 286.  Polarization calibration failed for unknown reasons
on 4 February 1989 and 24 April 1998 and these data were excluded
from further analysis.  The parallactic angle
coverage at 8.4 GHz was too limited to permit accurate leakage
term calibration.
Amplitude gains and leakage terms were transferred from J1743-0350
to the other sources.
These sources were then phase-only self-calibrated and imaged in
Stokes I, Q, U and V at 4.8 GHz and Stokes I and V at 8.4 GHz.  
Fluxes were determined by fitting a beam-sized
Gaussian at the image center.  All self-calibration and imaging
were performed using baselines longer than 100 $k\lambda$ at 4.8 GHz
and 150 $k\lambda$ at 8.4 GHz in order to exclude contamination from
extended structure.

The results are summarized Figures~1 to 9 and 
Table~\ref{tab:meancp}.
Figures~1 and
3 give the measured fractional CP for each source
at 4.8 and 8.4 GHz, respectively.  The plotted errors are the results of
thermal noise.  We see
clearly in these that there is an apparent variation 
common to all sources.  This indicates that some component of the 
measured signal is in fact due to instrumental error or due to 
variable CP in the amplitude calibration source,
J1743-0350.  We can estimate the degree of this variation with the
assumption that the sources W56, W109 and J1751-2523 are unpolarized.
The source GC 441 is probably also unpolarized but the limits on
fractional polarization in this source are too high because of its
low flux density.  We estimate the correction as the mean of the 
fractional CP for W56, W109 and J1751-2523.  This correction is
subtracted from the raw fractional CP of each source (Figures~2 and 4).  
The corrections are shown in Figure~5.
The application of the correction flattens the
light curve in fractional CP
for Sgr~A*.   It also reduces the scatter
between results for Sgr~A* in 6 of 8 epochs at 4.8 GHz (Figure~5).
We estimate the error with the rms scatter, which is on the
order of 0.1\%.  This is consistent with the theoretical
estimate of the error from our analysis in \S 4.
The results of April 1998 published in 
\citet{1999ApJ...523L..29B}
are more negative by $\sim 0.1\%$.

Figures~6 and 7 show the total intensity at 4.8
and 8.4 GHz, respectively.  
Figures~8 and 9 show Stokes Q and U at 4.8 GHz.
Plotted errors are thermal noise.
Sensitivity limits for Q and U are set at 0.1\% by
variations in the polarization leakage terms
\citep{1992VLASci...163}.

\section{VLA Observations in 1999}

Sgr~A* was observed by the VLA in the A array on 13 occasions in
1999.  These observations were spaced roughly by one week covering
the time range 23 June 1999 to 21 September 1999.  The shortest and
longest 
gaps between observations were 5 and 18 days, respectively.  
Each observation was approximately 4
hours in length.

On all but one day, observations were made at 1.4, 4.8, 8.4 and 15 GHz.
On 30 June 1999, observations were not made at 15 GHz.  All observations
were made with 50 MHz bandwidth in left (LCP) and right (RCP) circular
polarization.

Observing was divided into 5
blocks. Each block 
consisted of a 1.4 GHz segment and a 4.8, 8.4 and 15 GHz
segment.  Each segment began with a pointing scan on J1733-1304
at 1.4 GHz or 8.4 GHz.  Separate pointing scans were necessary because
of the uncertain alignment of the 1.4 GHz feed with the higher
frequency feeds.  The segments continued with observations
of J1733-1304, Sgr~A*, J1751-2523 and J1744-3116 at each frequency.
The higher frequency  feeds are collimated with each other.
J1751-2523 was not observed at 15 GHz due to its steep spectrum.

{\it A priori} amplitude calibration was performed with the source
3C 286 at all four frequencies.  Self-calibration on J1733-1304 
determined antenna-based amplitude gains.  These gains were
transferred to the other sources.  LP calibration was performed
with J1733-1304.  Electric vector position angle calibration
was performed using 3C 286.  All program sources were
phase-only self-calibrated and then imaged in Stokes I, Q, U  and V.
Fluxes were determined by fitting a beam-sized
Gaussian at the image center.  All self-calibration and imaging
were performed using baselines longer than 100 $k\lambda$ at frequencies
higher than 4.8 GHz
and baselines
longer than 30 $k\lambda$ at 1.4 GHz in order to exclude contamination from
extended structure.  Data from 10 August 1999 were corrupt
and not used.

We tabulate the mean total intensity and fractional CP
in Tables~\ref{tab:meani} and \ref{tab:meancp}.
We show in Figures~10 and 11 the results.  Figure~10 shows Stokes
V at all four frequencies for the three target sources, Sgr~A*,
J1751-2523 and J1744-3116.  Figure~11 shows the total intensity evolution
for the three target sources.  We do not plot LP
results but they clearly show no LP for Sgr A*
at all four frequencies at a level of $\sim 0.1\%$  on all dates. 
LP measured
for J1751-2523, J1744-3116 and J1733-1304 were slowly
variable.

\section{Error Analysis of VLA CP Measurements}

The VLA is equipped with RCP and LCP receivers.  Stokes V is measured
as the difference between the correlated parallel hands in polarization,
i.e. $V=1/2(RR - LL)$.  
Errors in CP measurements with the VLA have
seven origins:  one, thermal noise; two, gain errors; three, 
 beam squint; four, second-order leakage corrections; 
five, unknown calibrator polarization; 
six, background noise; and seven, radio frequency interference.
Primarily,
the first four of these are stochastic and the last three are 
systematic.
Our stochastic noise model is parametrized by $N$, the number of
2.5 minute scans, and $N_a$, the number of antennas 
(Table~\ref{tab:vnoise}).  For the monitoring
observations, $N=5 \pm 1$.  For the archive data, $N \approx 12$
at 4.8 GHz and $N\approx 4$ at 8.4 GHz.
We discuss each of the noise sources individually and then compare the
predictions of our model to the actual measurements.  
The agreement is good.  

\subsection{Thermal Noise}

Receiver noise introduces an error that is inversely
proportional to $\sqrt{N_a(N_a-1)BN}$, where $B=50$ MHz is the observing
bandwidth.  
For $N=5$ at 8.4 GHz, the error for the VLA is 40 $\mu {\rm Jy}$,
or 0.005\% for Sgr~A*.
This is an insignificant source of error for our VLA
results.

\subsection{Gain Errors}

Calibration can introduce a significant error in CP
measurements.  This can occur in two ways:  one, the amplitude
calibration source is too weak for accurate gain measurement;
two, the amplitude gains determined for the calibrator are not
the same as the gain on the target source.  Any errors in our
observations are due to the latter.  The primary amplitude calibrators,
J1743-0350 and J1733-1304, have flux densities on the order of 4 Jy, implying
that the SNR of amplitude calibration is on the order of $10^5$
for a 2.5-minute scan.  

We cannot easily separate the effects of errors from gain variation
with sky angle
and errors from gain variation with time.  
However, both errors will likely
have a similar signature in RCP and LCP, reducing
their effect significantly.
Furthermore,
the gain curves for VLA antennas
are known to be flat at these frequencies, although to what degree
is uncertain.  
Significant gain errors in angle would appear
as a systematic shift in the flux densities of the target sources, which
is not seen.  
We can estimate the combined effect of gain variations in time and
angle by measuring the variation of the gain solutions.
We find from our solutions
that the antenna gain variations are typically 0.3\%.  The
total error is inversely proportional to $\sqrt{N_aN}$, which
is on the order of 0.03\%.

\subsection{Beam Squint}

Beam squint is the result of offset RCP and LCP receivers in
the VLA antennas 
\citep{1991ATCA...39.3..0.15}.  
This introduces a false CP
off the beam axis.  Pointing errors can, therefore, lead to false 
CP.  In Figure~12, we show the false CP induced due to beam squint
for a single antenna and for the array, for pointing errors of
$10^{\prime\prime}$ (uncorrected pointing) and $2^{\prime\prime}$ (best case of
reference pointing).  The error due to beam squint scales inversely
with $\sqrt{N_aN}$, if pointing errors are uncorrelated.  Rejection
of antennas with large amplitude variations is an important
step before imaging for reducing the effect of beam squint.  A small
number of antennas with known pointing problems were routinely
rejected.

\subsection{Second-Order Leakage Terms}

Perfect receivers are sensitive to only a single polarization.  Real
receivers are sensitive to both LCP and RCP.  A complete treatment of
this polarization leakage leads to an identification of second-order
terms that affect Stokes V 
\citep[e.g.,][]{1994ApJ...427..718R}.
These terms 
are proportional to $|D|^2I$ and $DP$, where $I$ is the total intensity,
$P$ is the LP intensity and $D$ is the leakage
fraction.  $|D|$ is on the order of a few percent for the VLA.  
For the observed sources, $P/I \sim 1\%$.  For Sgr~A*, $P/I < 0.1\%$.  
The sum of these  terms contributes approximately 0.03\%.

\subsection{Unknown Calibrator Polarization}

This is a systematic error that is introduced in the calibration
step.  For self-calibration, we assume that the calibration
source is unpolarized.  The target source polarization will then
be offset from its true value by the calibrator polarization.
A strongly polarized calibrator will be detectable as a systematic
offset in the target sources.  This may have played a role in 
the correction factor applied to the archive data.  We show
below absolutely determined CP for 
our calibrator sources from ATCA.  These are all on the order of 0.1\%
or less.
This is probably the dominant source of systematic error in the
VLA measurements.

\subsection{Background Noise}

The strong background in the Galactic Center can introduce false CP.
This is a systematic error that is due to nonlinearity in the 
amplifier chains.  These are known to be linear to better
than 1\%.  This effect is greatest at low frequencies, where
the background is brightest.  At 4.8 GHz, the shift in system
temperature between Sgr A* and a calibrator source is from 35
to 25 K.  This implies an error $\mylesssim 0.3\%$ per polarization
per antenna.  Averaging, we find a total systematic offset
of $\mylesssim 0.04\%$.  The sign of this offset is unknown.

\subsection{Interference}

Interference can also introduce errors in CP.  Many satellite 
and terrestrial beacons
radiate in only a single circular polarization.  
Many of our observations are not made fully in 
protected radio bands.
Interference as a source of
CP can be identified if the CP for several nearby sources is similar
or if the CP is time-variable.  Our consistent detections
for Sgr A* at multiple frequencies
and non-detections for calibrator sources argue 
against interference errors.

\subsection{Comparison with Measured Errors}

We can estimate the actual error in CP by measuring
variations in the CP observed for the calibrator sources.
The best estimate of the noise in an individual measurement
is the rms difference between two consecutive measurements.
In an evenly sampled data set this is the structure function of CP
evaluated on the shortest sampling time scale.
We summarize in Table~\ref{tab:sigcp} these values for
Sgr A*, J1744-3116 and J1751-2523.  The variations for 
J1751-2523 are the lowest and, therefore, the most
indicative of the noise level at 1.4, 4.8 and 8.4 GHz.  
The measured values are within 50\% of the estimated
values (Table~\ref{tab:vnoise}), confirming our error model.  At 15 GHz, the 
measured variations for J1744-3116 are 0.14\%, significantly
higher than estimated.  There is probably a substantial
component from intrinsic variation in J1744-3116,
which is apparent at the 0.10\% level at the lower
frequencies.  This implies that the true error at
15 GHz is on the order of 0.10\%.

\section{ATCA Observations in 1999}

We observed Sgr A* with ATCA in the Spring and Fall of 1999.  These
observations are an important complement to the VLA observations
because they have an entirely different set of systematic
errors associated with them.  Most importantly, the ATCA receives
orthogonal linear polarizations.  The CP is
formed from the sum of correlated cross-hand visibilities rather
than from the difference of correlated parallel-hand visibilities.
The result is insensitive to amplitude calibration, beam squint,
and calibrator polarization.  Furthermore, the differences in antenna
size and shape,  array configuration  and location
provide a test against effects of background, second-order leakage
term and interference.  
Additionally, separate calibrators and analysis techniques
were used.  The errors are dominated by thermal effects and leakage
term measurements.

Standard polarization observation and reduction techniques are
described in 
\citet{1991ATCA...39.3..0.15}.
These methods have been shown to produce 
errors in CP $\sim 0.01\%$.
All analysis was performed with the MIRIAD
software \citet{1995adass...4..433S}.  We observed the source B1934-638 to
calibrate amplitudes and to determine leakage terms.  
The compact source J1820-2528 was observed to determine 
the time-dependent phase solutions.  These were applied to Sgr A*
and other calibrators.  Observations were performed in the extended 
6A and 6D configurations of the ATCA, which have a maximum
baseline of 6 km.  Only baselines longer than 30 $k\lambda$
were used in analysis of Sgr A*.  This smaller limit on the
baseline length is necessary because of the more compact
configuration of the ATCA.

Observations in the boreal Spring of 1999 were conducted 
on 11 May, 22 May and 02 June.  These observations were each
12 h tracks on Sgr A*.  Sky frequencies were centered at
4.8 GHz and 8.6 GHz with 128 MHz of bandwidth in two
polarizations at each frequency.  Several calibrators
were observed in addition to  J1820-2528.
Limited
observations of J1751-2523, J1733-1304, J1743-038,
W56, W109 and GC 441 were also made.

Observations in the boreal Fall of 1999 were made on 
8 dates between 11 September and 04 October.  Typical
tracks were 6h in duration.  The sky frequencies were
8.512 GHz and 8.640 GHz with 128 MHz bandwidth at
each frequency.  Again, J1820-2528 was used as the main 
phase calibrator.  J1744-3116 and J1751-2523 were also routinely observed

We plot in Figures~10 and 11 
the results of the ATCA observations for Sgr A*,
J1744-3116 and J1751-2523.  Mean CP is listed in Table~\ref{tab:meancp}
for all sources.  We also produced LP results which are consistent
with the VLA results.

We have a few ATCA measurements of the CP in our VLA calibrators
J1733-1304 and J1743-038.
These low values confirm that the CP in these sources does not
strongly bias the VLA results.

\section{Timescales Greater than 1 Year}

{}From the VLA archive data we see that the CP
of Sgr~A* is roughly constant over 18 y.  The mean is
$m_c=-0.31 \pm 0.13\%$ at 4.8 GHz and $m_c=-0.36 \pm 0.10\%$
at 8.4 GHz.  The errors are measures of scatter. 
Because of our reduction methods, we cannot make
any statements about variability within an epoch.  However,
we can estimate whether there is variability between epochs
if we estimate the error in $m_c$ for each epoch as the scatter
in the measurements added in quadrature with a systematic error
of 0.05\%.  Under the assumption of constant $m_c$, the reduced $\chi^2$
is 1.6 at 4.8 GHz and 1.1 at 8.4 GHz.  
This is consistent with constant $m_c$ 
for the typical error of 0.1\% 
on timescales longer than 1 y and less than 18 y.
The mean spectral index is slightly
inverted.  We find
for $m_c \propto \nu^\alpha$, $\alpha=0.2 \pm 0.7$.

The constant fractional CP occurs simultaneously with
a slow change in the total intensity.  The total intensity
decreases by a factor of $\sim 2$ in the 18 y period.
We find no evidence for LP in Sgr~A* at 4.8 GHz
above 0.1\% over 18 y.

\section{Timescales Between 1 Day and 1 Year}

We can use the VLA and ATCA observations to constrain the degree
of variability of CP in Sgr A* on
timescales between 3 and 100 days.  We can also use the VLA
data to characterize the spectral evolution of variability from
1.4 to 15 GHz on these timescales.
There is good agreement between the results from the VLA and ATCA
at 4.8 and 8.4 GHz.  

First, we discuss the nature of total intensity variability
in Sgr A*. 
We can construct a structure function 
from the VLA and ATCA
data.  The structure function is defined as 
\begin{equation}
D^2 (\tau)= {1 \over T}\int^T_0 dt \left[ S (t+\tau) - S (t) \right]^2.
\end{equation}
In order to compute this continuous function for our
discretely and irregularly sampled data, we calculate the 
difference for each pair of points and average them in
bins.  We plot for both sources the structure function of total intensity
(Figure~\ref{fig:isfun}).
Variability increases strongly with frequency.
We also see that 
the structure function rises with $\tau$ on timescales of 3 to 100 days
at all frequencies.  At 4.8 GHz, the structure function continues
to increases out to $>1000$ days.
These results echo
the earlier conclusions of \citet{1992rbag.work..295Z}, 
\citet{1994nngl.conf..403B} and \citet{1999cpg..conf..113F}.

From the CP data,
we draw several conclusions.  One, the mean fractional CP spectrum
is inverted.  Two, the degree
of variability increases with frequency.  Three, the time scale
for variability is as short as a few days to a week.  Four, 
the largest change in fractional CP was 
coincident with the largest total intensity variation.

We plot in Figure~\ref{fig:meanvar} the mean spectrum 
and degree of variability 
of fractional CP in Sgr A*.  The ATCA, VLA and VLA archive results are in good
agreement at 4.8 and 8.4 GHz (Table~\ref{tab:meancp}).  
The mean spectrum in the Summer of 1999
is inverted with $\alpha=0.5 \pm 0.2$.  
This figure demonstrates
the increasing degree of variability with frequency.
The degree of variability at 1.4 GHz is
consistent with the noise, while the degree of variability
at 15 GHz is 
three times greater than the degree of variability
for J1744-3116 (Table~\ref{tab:meancp}).

We derive the structure function for the fractional CP in a similar 
manner to that of the total intensity.
The structure functions for Sgr A* and J1744-3116 are plotted in 
Figure~\ref{fig:sfun}.  These structure functions demonstrate
a number of our conclusions.  One, the increasing degree 
of CP variability with frequency is again apparent in the VLA
data.  Two, the degree of CP
variability at 4.8 GHz is 
constant on timescales of 10 to 1000 days.  The degrees
of CP variability at 1.4, 8.4 and 15 GHz are constant on
timescales of 10 to 100 days.  Three,
episodic {CP} variability has a substantial impact.  The ATCA {CP}
structure function at 8.4 GHz is substantially lower than
that of the VLA at 8.4 GHz.  Presumably, this is due
to one or two major flares during the VLA monitoring and
the absence of such events in the ATCA time periods.
Four, the ATCA data indicate that there is a sharp 
decrease in the degree of  {CP} variability on timescales less
than 10 days.  This may be true at all times or it 
may be a consequence of the lower overall variability
during the ATCA epochs.  The VLA data are not
sensitive to timescales much less than 7 days.  The
combined VLA and ATCA structure function at 8.4 GHz
(not shown) is consistent with the structure function
converging to the 4.8 GHz value on long timescales.
Finally, the structure functions for Sgr A* are substantially
greater than that of J1744-3116, emphasizing the reality
of the variability.
Results for
J1751-2523 are qualitatively similar but
several times weaker than those of J1744-3116.
There is also good agreement between the ATCA
and VLA results at 8.4 GHz for J1751-2523.

The VLA light curves indicate that 10 days 
is a characteristic timescale for
the rise and decay of flares in total intensity and CP
(Figures~10, 11 and \ref{fig:spectrum}).
Between day 210 and 217, there is a jump in $m_c$ from
-0.1\% to -1.1\% at 15 GHz.  This is followed by 
slow decline in $m_c$ over 20 d before another, smaller jump
in $m_c$ appears.  These events are
not correlated with any change in $m_c$ for J1744-3116.
The change between day 210 and 217 is coincident
with a significant flare in the total intensity of Sgr A*.
The increase in $m_c$ between day 240 and 246 does not
appear to have a correlated change in total intensity
but it is of a smaller magnitude.

With the sampling available, there is little evidence for
a time delay between flares at 15 GHz and 8.4 GHz.  The flare 
on day 217 
is not apparent at frequencies below 8.4 GHz.  However,
the flare at day 246 does appear at 4.8 GHz, although
it appears to be of lesser magnitude at 15 GHz than
the previous flare.

\section{Timescales Less than 1 Day}

We analyzed the 1999 VLA data on timescales of less than 1 day,
as well.  We computed the Stokes I and V terms for 30 second
averages.  The results clearly indicate episodic activity.
On some days, there is no evidence for variability in either
term.  However, at the time of the peak flux density (day 217) we see 
a $\sim 20\%$ change in the total flux density over a period of two
hours at 15 GHz (Figure~\ref{fig:short0805x4}).  
Similar, but less prominent changes are
apparent at 8.4 and 4.8 GHz.  The flux density at these lower
frequencies does not clearly saturate in the way that it does
at 15 GHz.  No change is apparent at 1.4 GHz.  A factor of
two change in Stokes V is apparent at 15 GHz over the same time
range with $\chi^2_{\nu}=3.7$ (Figure~\ref{fig:short0805x4v}).  
There is a delay between the
onset of variability in Stokes I and Stokes V of at least 1
hour.  No significant change is
apparent at any other frequency in Stokes V due to a lack of
sensitivity. 

Short-term variability was convincingly seen ($\Delta I > 100$ mJy at 15 GHz)
in only one other epoch.  On day 204, the 15 GHz flux density
increased by 150 mJy in two hours.  The 8.4 GHz flux density also increased
but by a lesser amount and no variability was detected at the lower
frequency.  There is no evidence for fluctuations in 
CP on this date.

We computed mean structure functions from the VLA data at a timescale
of 1 hour for Sgr A* and J1744-3116 in total intensity and fractional CP
(Figures~\ref{fig:isfun} and \ref{fig:sfun}).  In total intensity, Sgr A*
is significantly more variable than J1744-3116.  The degree of variability
increases with frequency as it does on longer timescales.  The structure
function increases roughly as $\tau^{1/2}$ from 1 hour to 100 days.  That
power law holds to $\sim 3000$ days at 4.8 GHz.  The power law index is
slightly larger than that found by \citet{1994nngl.conf..403B}.
In fractional CP
the structure function on short timescales is dominated by noise.  The
structure function for Sgr A* and J1744-3116 are similar at 1 hour.  However,
the structure for Sgr A* does increase between 1 hour and 10 days.
The high fractional CP structure function at 1.4 GHz is due to the
complex extended structure in the Galactic Center, which is not
well-modeled on short intervals.

\section{Discussion}

We discussed in detail in \citet{1999ApJ...523L..29B} some of
the difficulties in producing high CP and low LP at centimeter
wavelengths with a synchrotron source. Interstellar propagation
effects predict a very steep-spectrum behavior and are not very
efficient \citep{2000ApJ...545..798M}, so that they cannot be invoked as
a major contibutor to the overall CP of Sgr A*.  We conclude that
the CP we measure is intrinsic to Sgr A*. The responsible mechanism
then is either circular polarization from gyrosynchrotron emission
\citep{1969ApJ...158..753R} or re-polarization/conversion of LP to CP
\citep{1977OISNP..89.....P} --- both processes are most efficient in
the presence of low-energy electrons with Lorentz factors of only a
few. The introduction of a high density region of non-relativistic
electrons, either in the accretion region for $r<0.01$ pc or in the
source itself, can also lead to high depolarization factors needed to
suppress the LP \citep{1999ApJ...521..582B,2000ApJ...545..842Q}.

Moreover, since the amplitude of total and polarized intensity variability is
a function of frequency \citep{2000ApJ...545L.117M} single
component, optically thin models \citep[e.g.,][]{1994A&A...286..431D}
where all frequencies are produced at the same location and
scale up and down together, cannot account for these properties.
Inhomogenous, multi-component models must be used. This fits with
the fact that conversion can produce very high degrees of CP in
inhomogeneous sources with a range of spectral indices, including
inverted and flat spectra
\citep{1977ApJ...214..522J,1988ApJ...332..678J}.

One can split the total intensity spectrum of Sgr A* in two or three
regimes.  The millimeter/submillimeter has
an excess of radiation compared to the cm-wave spectrum
\citep{1997ApJ...490L..77S,1998ApJ...499..731F}. 
This bump is most likely associated
with a very compact region near Sgr A*.  In addition, the cm-wave
spectrum of Sgr A* may also be separated in two regimes, above and
below $\sim10$ GHz, based on the
fact that the variability behaviour seems to change
\citep{1998ApJ...499..731F,2001ApJ...547L..29Z}.

We may be able to consider the CP spectrum as composed of 
two components, as well.  The low frequency component may be 
determined from the spectrum on day 210 (Figure~\ref{fig:spectrum}).  This 
component has a spectral peak near 4.8 GHz and an
index $\alpha\sim -0.5$ between 4.8 and 15 GHz.  This
component is apparently steady.  The high frequency
component is highly variable in the short term.
The spectrum on day 217 indicates that the spectrum
can be strongly inverted with $\alpha\approx 1.5$.
The apparent decay time for this component is
on the order of 20 days.  The decay time is less than
the decay time in total intensity, which is $\sim$
10 days.  

If the higher frequency CP component is associated with the
millimeter/submillimeter component, then the fractional
CP of that component is much higher
than the measured value.  In a typical ADAF model, the millimeter/submillimeter
component is modeled as thermal gyrosynchrotron emission.  At 
15 GHz, this thermal component contributes approximately 10\%
of the total intensity.  This ratio holds roughly for the jet model 
\citep{2000A&A...362..113F}
as well as for the low $\dot{M}$ model of 
\citep{2000ApJ...545..842Q}.
Thus, the maximum CP
at 15 GHz corresponds to $\sim -10\%$ of the total intensity
of the millimeter/submillimeter component.

The coincidence of the brightest CP flare with the
brightest total intensity flare suggests that the processes driving
these phenomena are fundamentally linked.   This flare occurred in
phase with the 106 day period in total intensity
\citep{2001ApJ...547L..29Z}, suggesting that the CP may also
be periodic.
Since our CP data are only weakly sensitive
to timescales of 100 days, we cannot confirm the periodicity in 
CP.  However, sharper definition of the
period phase and future CP observations
will test this 
relationship more rigorously.  
\citet{2001ApJ...547L..29Z}
found that the periodic 
variability of Sgr A* at 15 GHz has an amplitude 
on the order of 30\% of the total intensity.  If the CP
at 15 GHz is solely associated with this component, then it is
$\sim -4\%$ polarized at its maximum.

An extrapolation of the high frequency CP spectrum at its peak flare
state indicates that $m_c$ may exceed 10\% at 100 GHz.  A single observation
at 112 GHz found $m_c < 1.8\%$, indicating that if it does reach 10\%,
it does so episodically \citep{2001ApJ...555L.103B}.

The long-term stability of $m_c$ at 4.8 GHz combined with a factor-of-two
change in total intensity suggests a linear relationship between
CP and total intensity.  A propagation effect
such as birefringent scattering 
\citep{1999ApJ...523L..29B,2000ApJ...545..798M}
could produce this relationship although
it has difficulty accounting for the relatively flat spectrum
of the CP.  Simple synchrotron emission will produce
this effect, too.   This requires uniformity in the magnetic pole
and accretion conditions over 20 y.
Timescales for turbulence in the wind-driven accretion
region are on this order of magnitude 
\citep{1999ApJ...523..642C}.  This is a much more interesting
limit on the stability of the handedness of CP than 
the 20 y limit found for luminous AGN, since intrinsic timescales
typically scale with the mass of the black hole \citep{2001ApJ...556..113H}.
On the other hand, the short timescale variations in GRS 1915+105
probe a very different accretion regime 
\citep{2001MNRASprepfender}.

The power law index of the total intensity structure function also
emphasizes the differences between Sgr A* and high luminosity AGN.
\citet{1992ApJ...396..469H} found no sources with a power law index less than 0.7
on timescales of months to $\sim 10$ years.  These differences are not
only the result of mass differences (if timescales scale linearly
with mass).  Our structure function timescales of 1 hour to 100 days
would correspond to 1 month to 300 years.   Thus, we conclude that 
variability on short timescales is qualitatively different from
that of high luminosity AGN.

Variability in total intensity and in circular polarization 
appears to occur on a continuum of timescales from 1 hour
to $>$ 1000 days.  The 1-hour timescale is comparable to the timescale
for a $\sim 50$ times flare observed at X-ray wavelengths
\citep{2001Natur.413...45B}.  The substantially lower amplitude of the
radio variability probably implies an inverse-Compton origin for the
X-ray photons \citep{2001A&A...379L..13M}.  

We can estimate an upper limit to the size of Sgr A* at 15 GHz
given $\sim 20\%$ variability over 2 hours.  An upper limit to
the 
size of the outburst is then $c\times 2 {\rm\ h} / 0.2 \sim 70$ AU,
which is 8 mas at the Galactic Center.  This is larger than
the 15 GHz scattering size of Sgr A* (5.4 mas).  
The observed source size will be 10 mas.  If the source
does grow to this upper limit size, then it could be readily
detected with VLBI at 15 GHz and higher frequencies even if the
intrinsic size is proportional to the inverse frequency.
There have been no convincing VLBI observations to indicate a deviation
from the scattering size at any frequency, however.
One could also expect a change in in the centroid of emission 
that is comparable to the change in the source size.
Astrometric experiments in the Galactic
Center have achieved sub-mas accuracy \citep{1999ApJ...524..816R}.

The flux density of Sgr A* appears to decline on a timescale of 10 days.
Synchrotron cooling is sufficient only for an ADAF model, where the
cooling time at 15 GHz is 25 days.  The low density model of 
\citet{2000ApJ...545..842Q} and the jet model \citep{2000A&A...362..113F}
predict cooling times of $\sim 200$ days.  Alfven velocities for the 
ADAF and \citet{2000ApJ...545..842Q} model are on
the order of $0.1c R^{-0.25}$, where $R$ is the radius in units of the gravitational
radius.  The crossing time for a source 1 light-hour across would be about
1 day.  For a jet model, the Alfven velocity is $\sim 0.5c R^{1/7}$.
The crossing time would be about 1 hour, implying that continuous energization
of the emitting region is necessary.  This is not unreasonable given the 
substantial short-term variability seen at the peak of the flare.

We note also that the effects of scattering do not operate on a 10-d timescale.  
For a relative velocity $v=100 {\rm\ km\ s^{-1}}$ between Earth
and the scattering medium, we can calculate the diffractive
and refractive timescales at 15 GHz.  
The diffractive timescale is $\sim 3 {\rm\ s}$.  The refractive
timescale is $\sim 2 {\rm\ y}$.   

The similar LP and CP properties of M81* and Sgr A* suggest that 
there may be a class of low luminosity AGN with these polarization
properties \citep{2001ApJ...560L.123B}.  In addition to an absence
of LP and the presence of CP (with a potentially inverted spectrum),
the sources both show unusual inverted total intensity spectra.
Such total intensity
spectra are the result of inhomogeneous electron energy and magnetic
field distributions.

\section{Conclusions}

We have presented here VLA and ATCA observations of CP and LP in Sgr A* 
that span timescales
of 1 hour to 20 y  at frequencies between 1.4 and 15 GHz.  CP 
is clearly detected at all frequencies.  In the short-term,
it is variable with
an increasing degree of variability with frequency.  The average
spectrum is inverted.  In the long-term, the 4.8 GHz CP is
remarkably stable despite changes in total intensity.
LP is not detected at any frequency.

We suggest that the CP has two separate components
in frequency.  A low frequency component peaks around 5 GHz
and then decreases with $\alpha \sim -0.5$.  A high frequency
component is strongly variable.  At maximum, the spectrum
is inverted at 15 GHz with $\alpha\approx 1.5$ and $m_c\approx -1\%$.  
At minimum, the high frequency component disappears at the 0.1\%
level.  The coincidence of the maximum in CP with a total intensity
maximum suggests that these flaring events are related.   The
two component nature of the total intensity spectrum suggests that
the higher frequency component may be $\sim 10\%$ circularly polarized at
the flare peak.  

Future millimeter wavelength observations of CP in Sgr A* can
test this hypothesis.  These observations should be timed   
to test the relationship between flaring in total intensity and
in CP.  

Finally, we note that these observations demonstrate the importance
of CP for diagnosing the physics of compact radio sources.  CP
observations can constrain models for accretion, outflow and
interstellar wave propagation.  

\acknowledgements
The National Radio Astronomy
Observatory is a facility of the National Science Foundation operated under
cooperative agreement by Associated Universities, Inc.
The Australia Telescope Compact Array  is part of the Australia Telescope
which is funded by the Commonwealth of Australia for operation as a 
National Facility managed by CSIRO. 


\newpage

\plotone{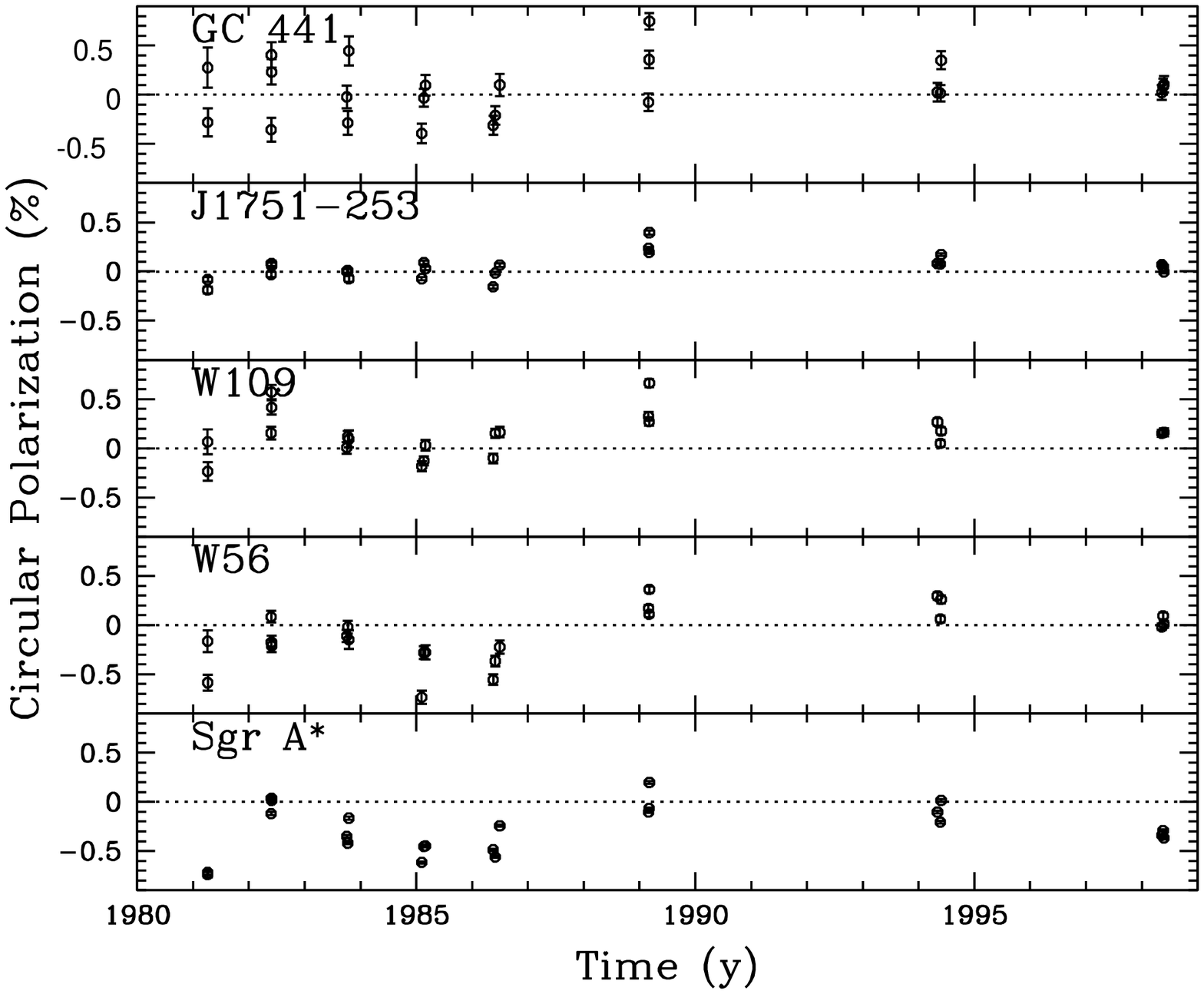}
\figcaption[f1.eps]{Raw fractional CP for five sources
with 4.8 GHz data taken from the VLA archives.  Error bars represent thermal
noise.}

\plotone{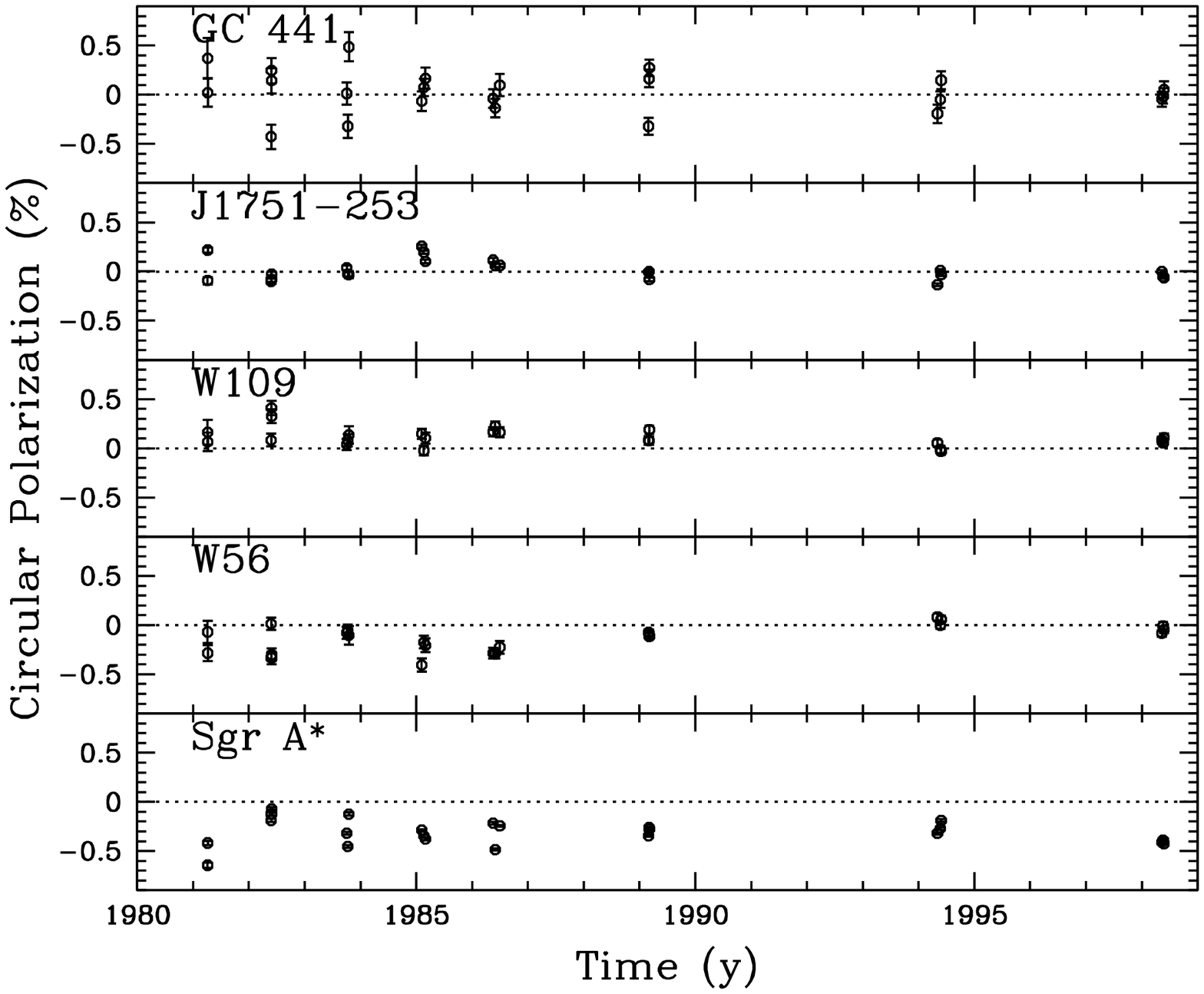}
\figcaption[f2.eps]{Corrected fractional CP for five sources
with 4.8 GHz data taken from the VLA archives}.

\plotone{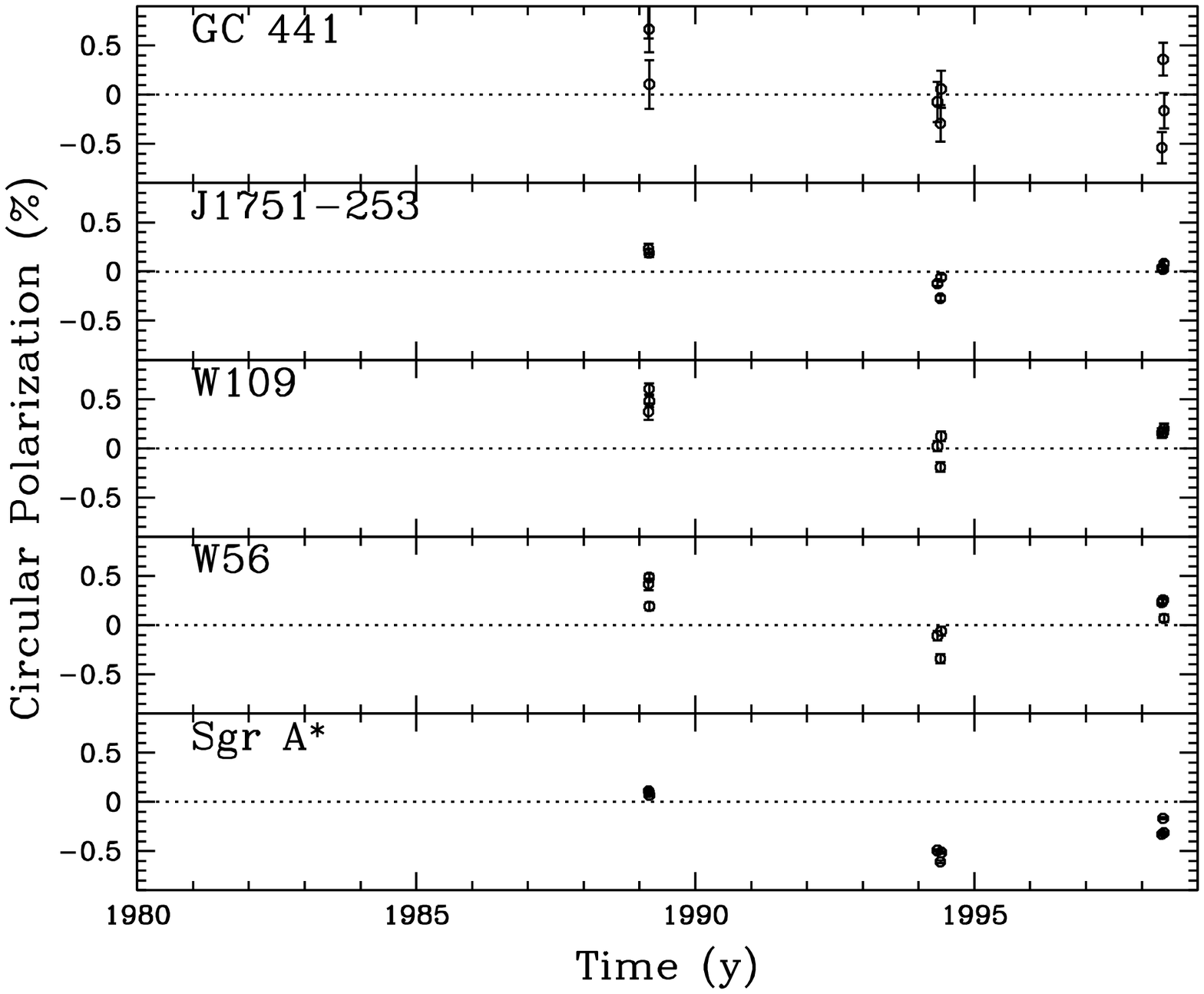}
\figcaption[f3.eps]{Raw fractional CP for five sources
with 8.4 GHz data taken from the VLA archives.  Error bars represent thermal
noise.}

\plotone{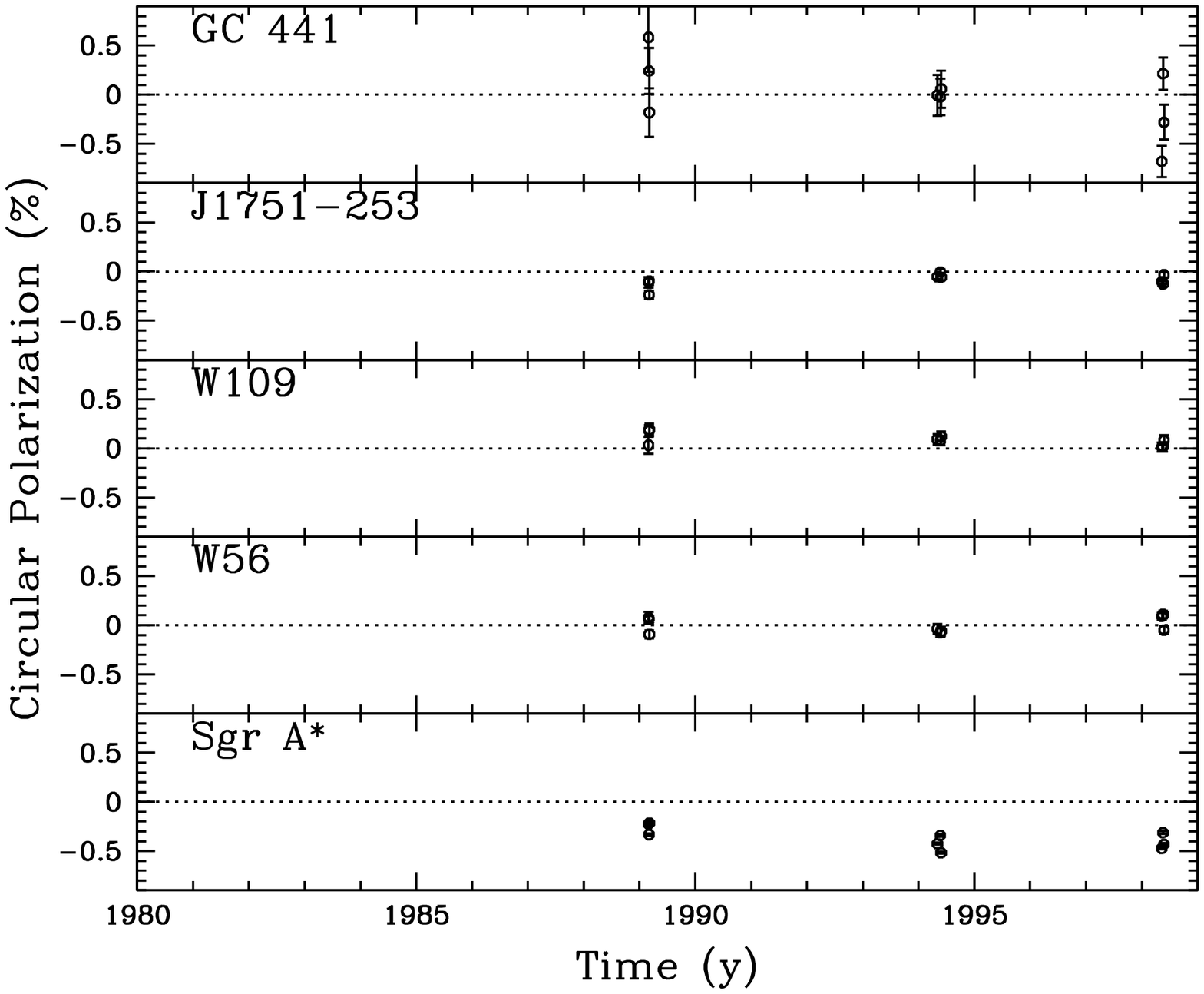}
\figcaption[f4.eps]{Corrected fractional CP for five sources
with 8.4 GHz data taken from the VLA archives}.

\plotone{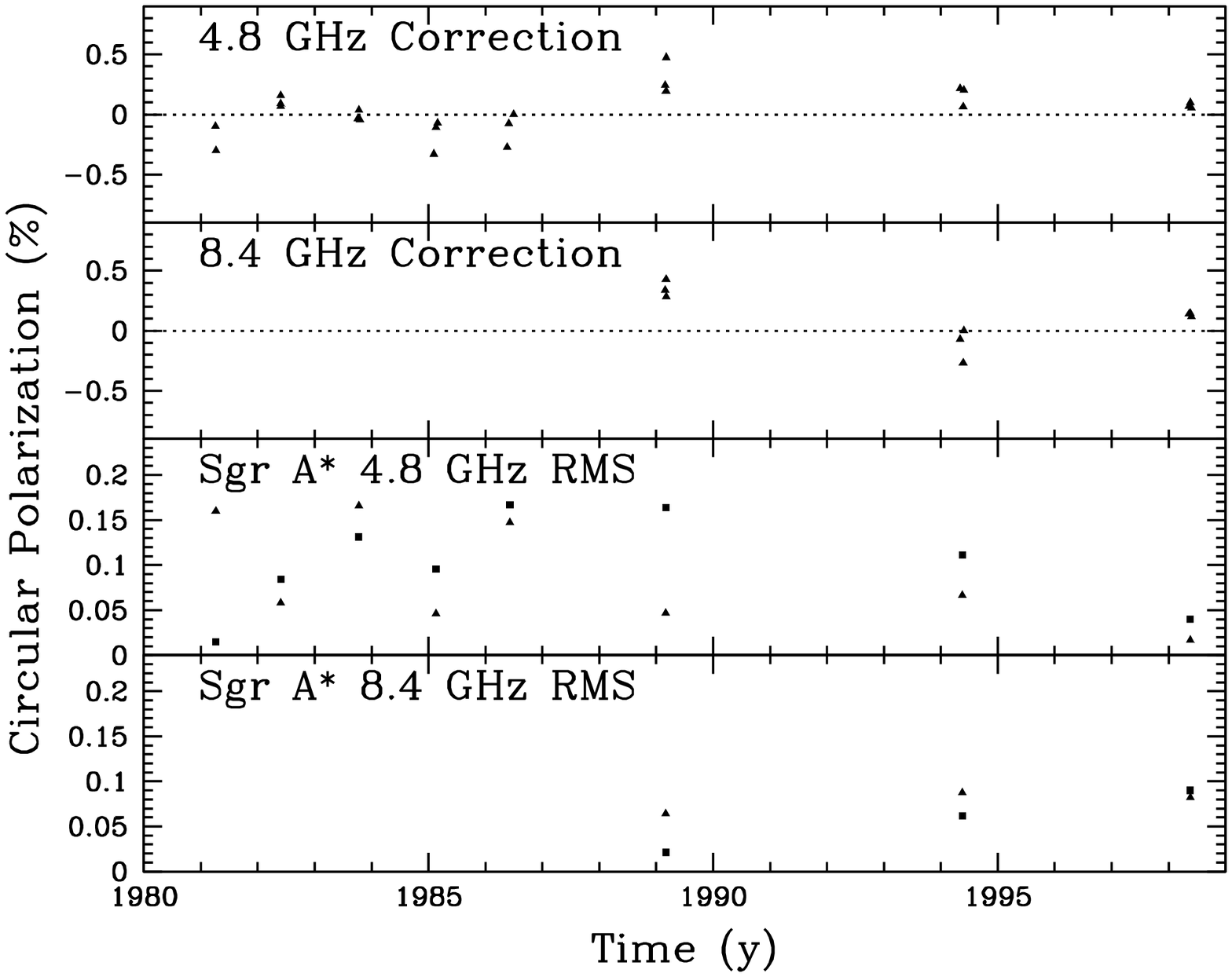}
\figcaption[f5.eps]{Correction terms for the raw fractional CP
at 4.8 GHz (top panel) and at 8.4 GHz (second from top panel) and
their effect on the rms scatter for Sgr~A*
within epochs at 4.8 GHz (second from
bottom panel) and at 8.4 GHz (bottom panel).  Triangles in the bottom
panels represent corrected data.  Squares represent uncorrected data.}

\plotone{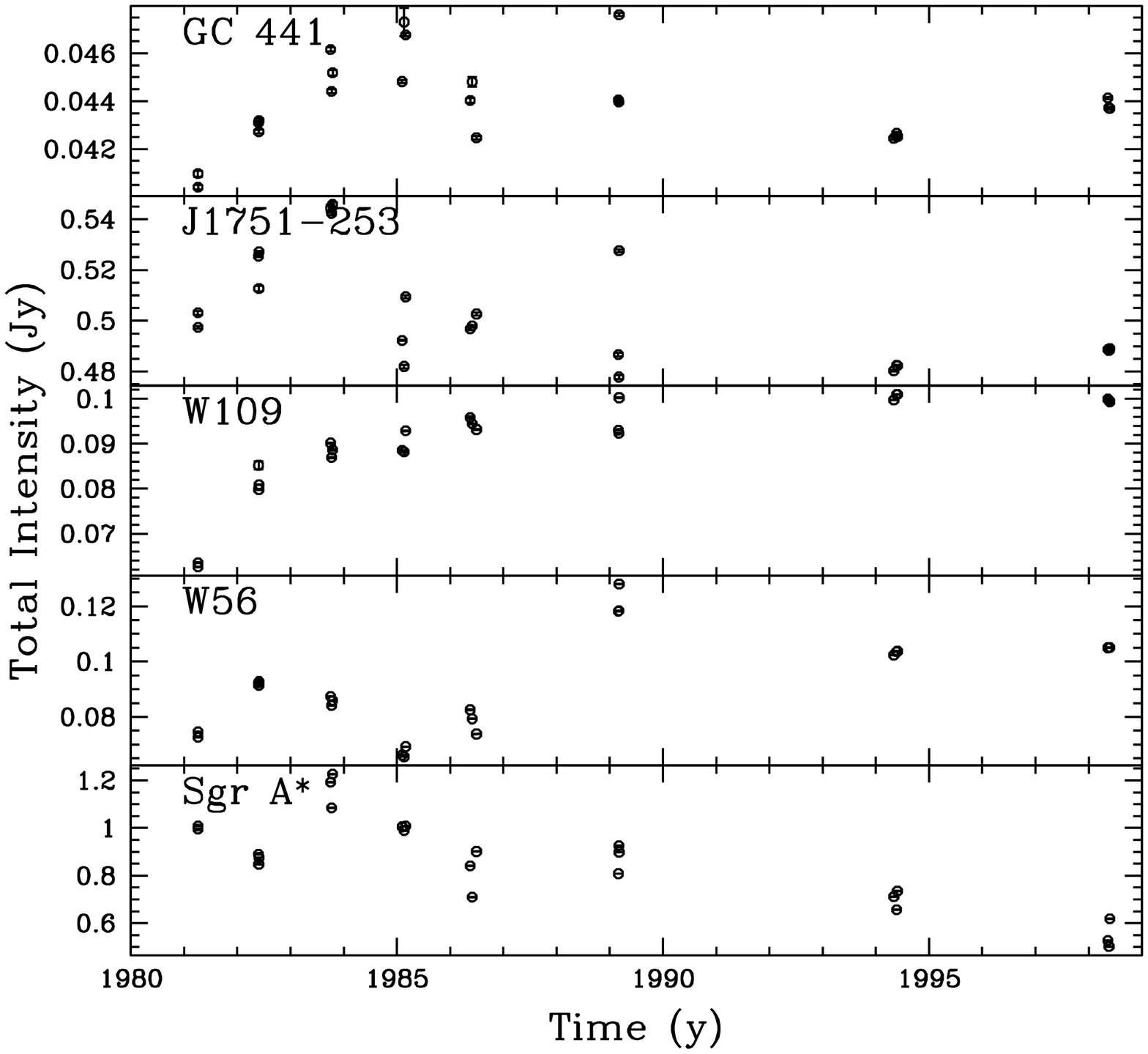}
\figcaption[f6.eps]{Total intensity for five sources with 4.8 GHz
data taken from the VLA archives.}

\plotone{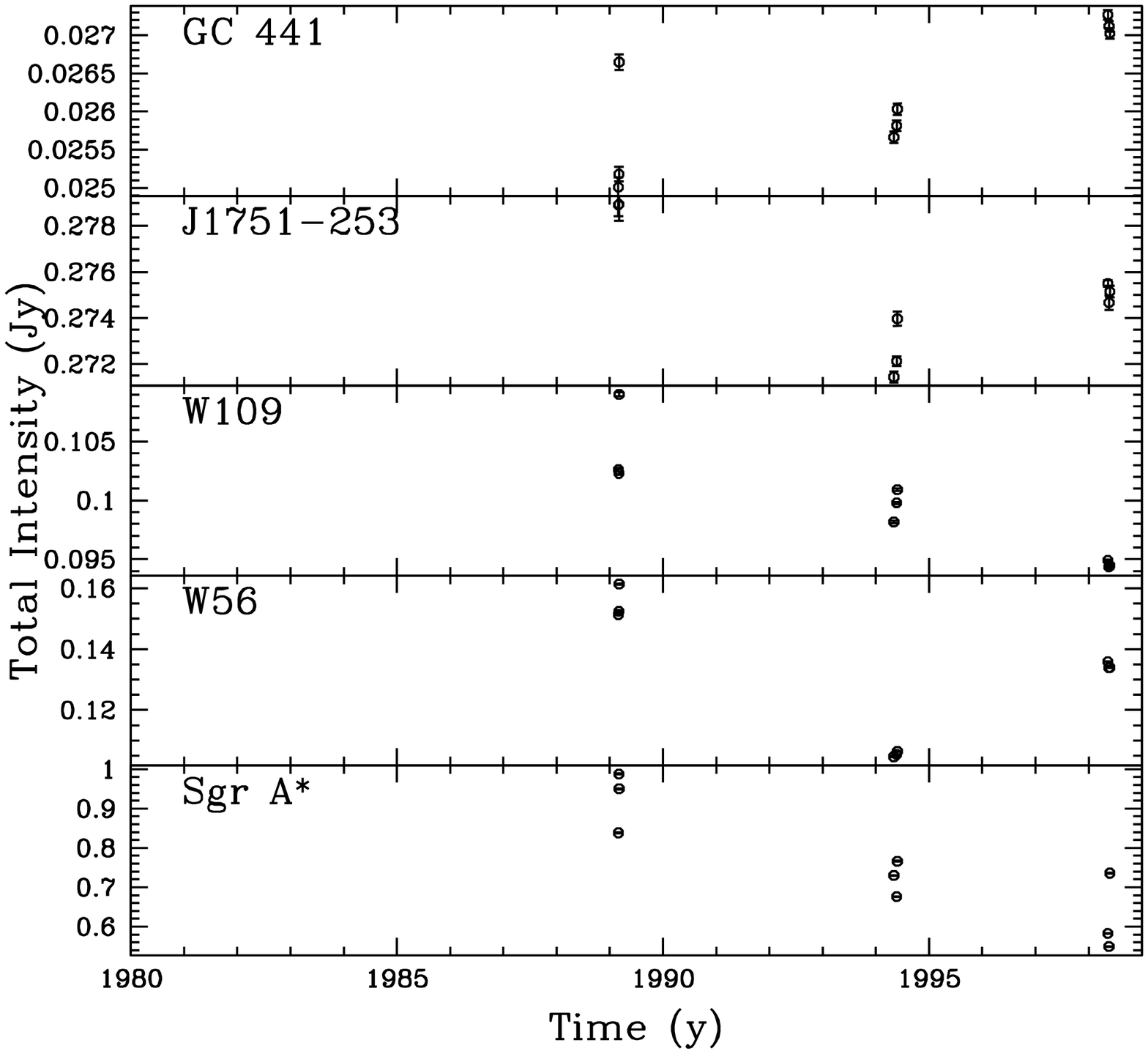}
\figcaption[f7.eps]{Total intensity for five sources with 8.4 GHz
data taken from the VLA archives.}

\plotone{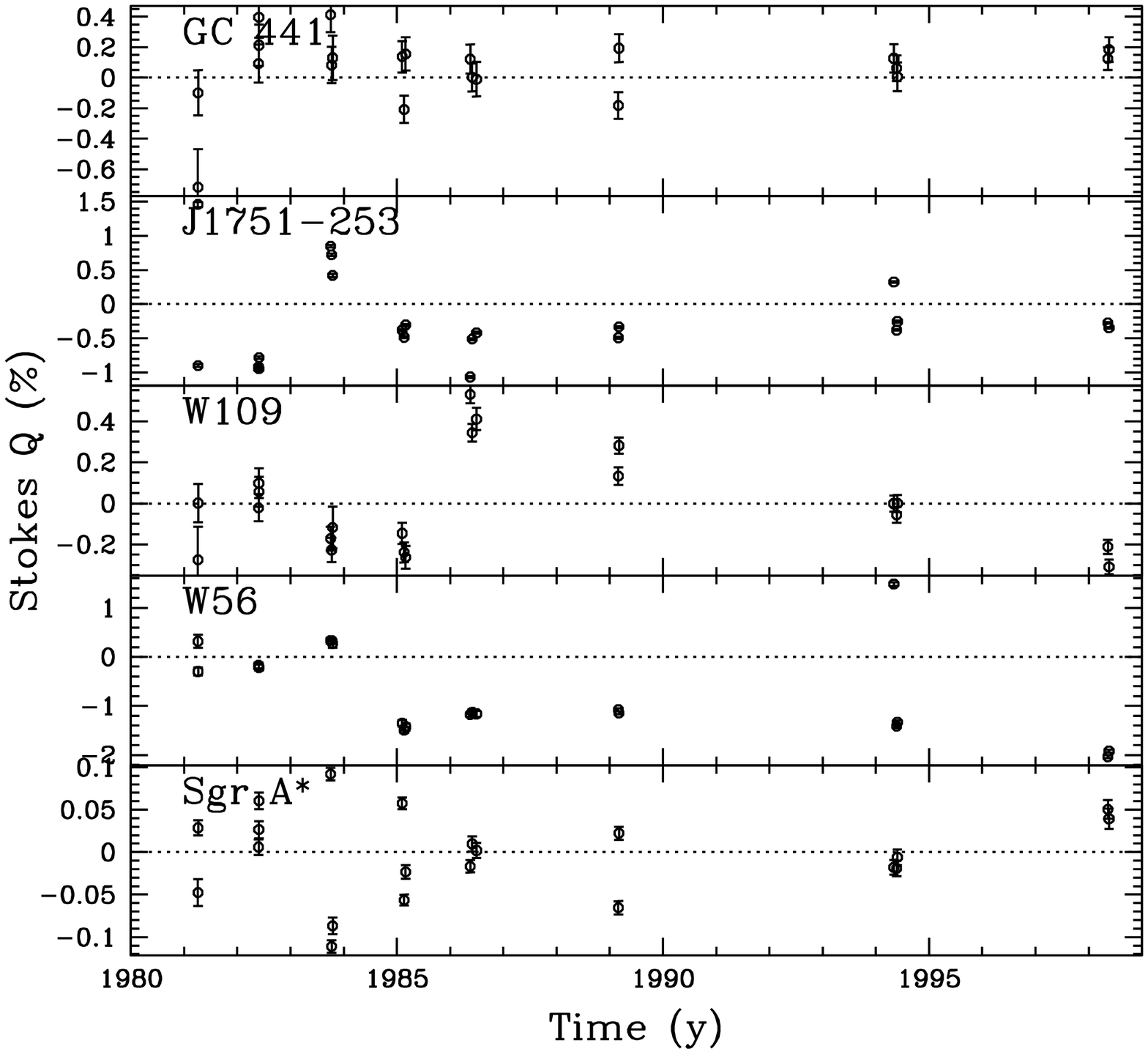}
\figcaption[f8.eps]{Fractional Stokes Q for five sources with 4.8 GHz
data taken from the VLA archives.}

\plotone{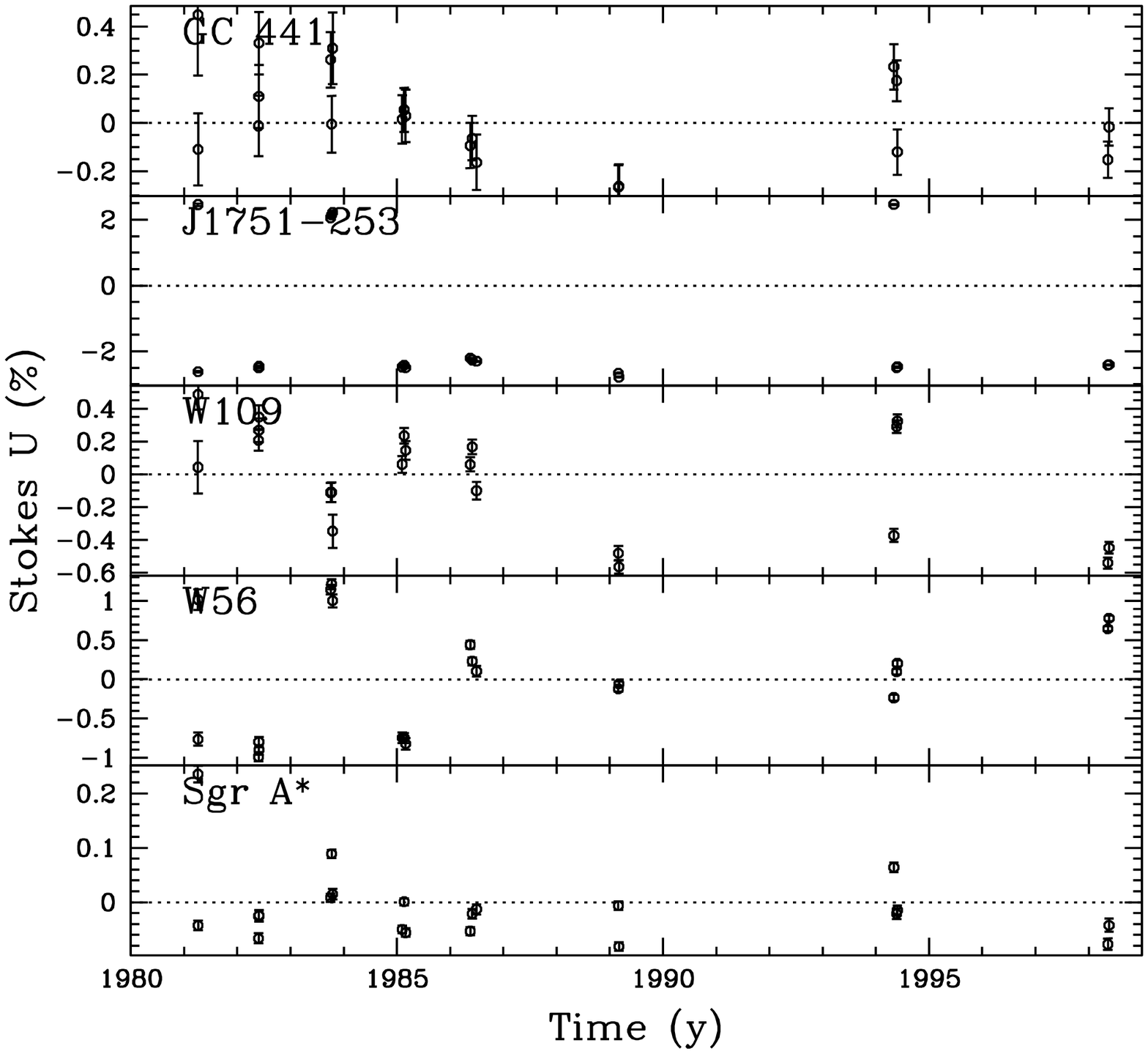}
\figcaption[f9.eps]{Fractional Stokes U for five sources with 4.8 GHz
data taken from the VLA archives.}

\plotone{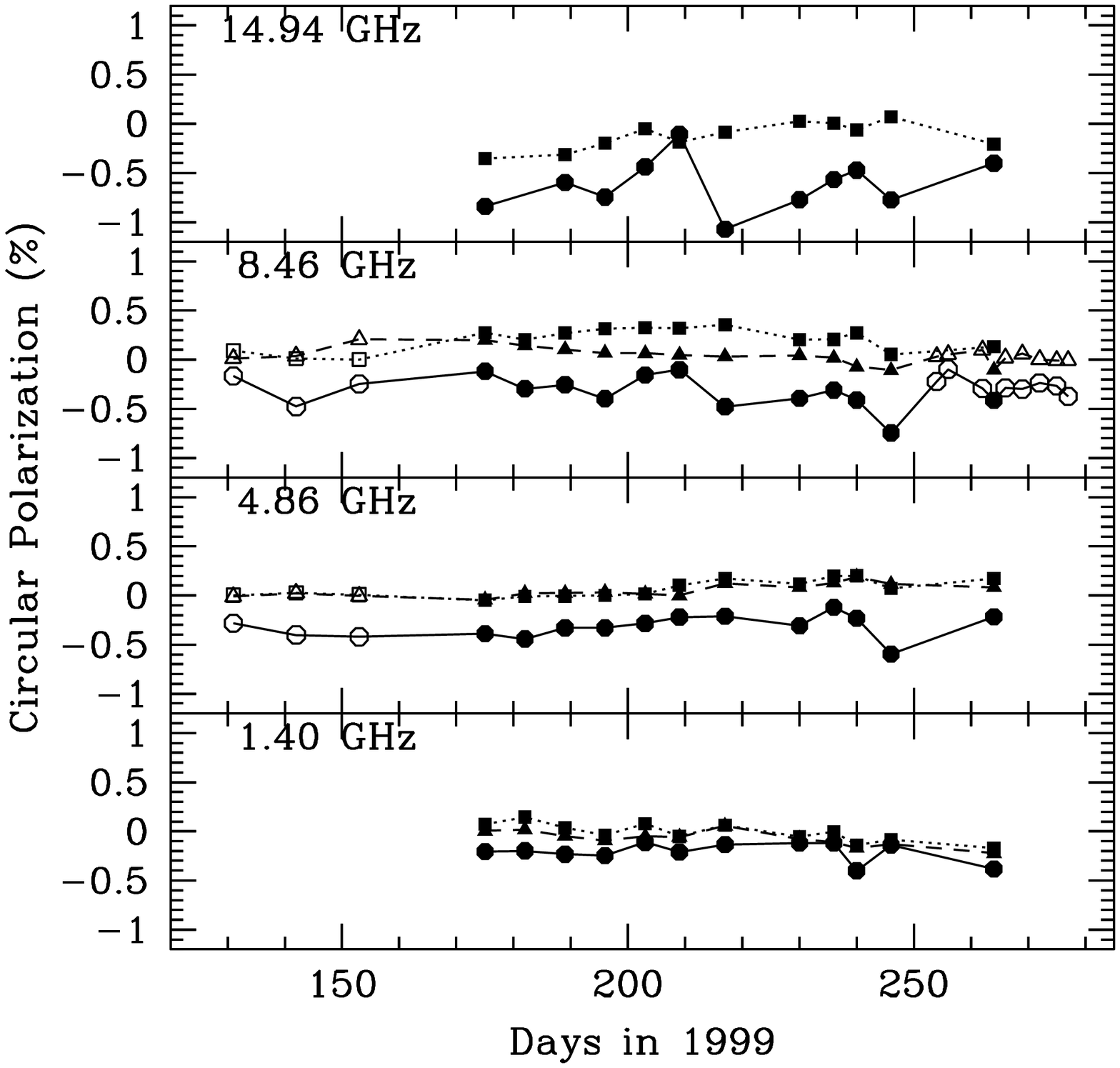}
\figcaption[f10.eps]{Fractional CP for three
sources at 1.4, 4.8, 8.4 and 14.9 GHz from the 1999 VLA and ATCA
monitoring campaigns.
Octagons and solid lines are for Sgr~A*, squares and short-dashed lines
are for J1744-3116, and triangles and long-dashed lines
are for J1751-2523.  Solid symbols are for the VLA and open symbols
are for ATCA.}

\plotone{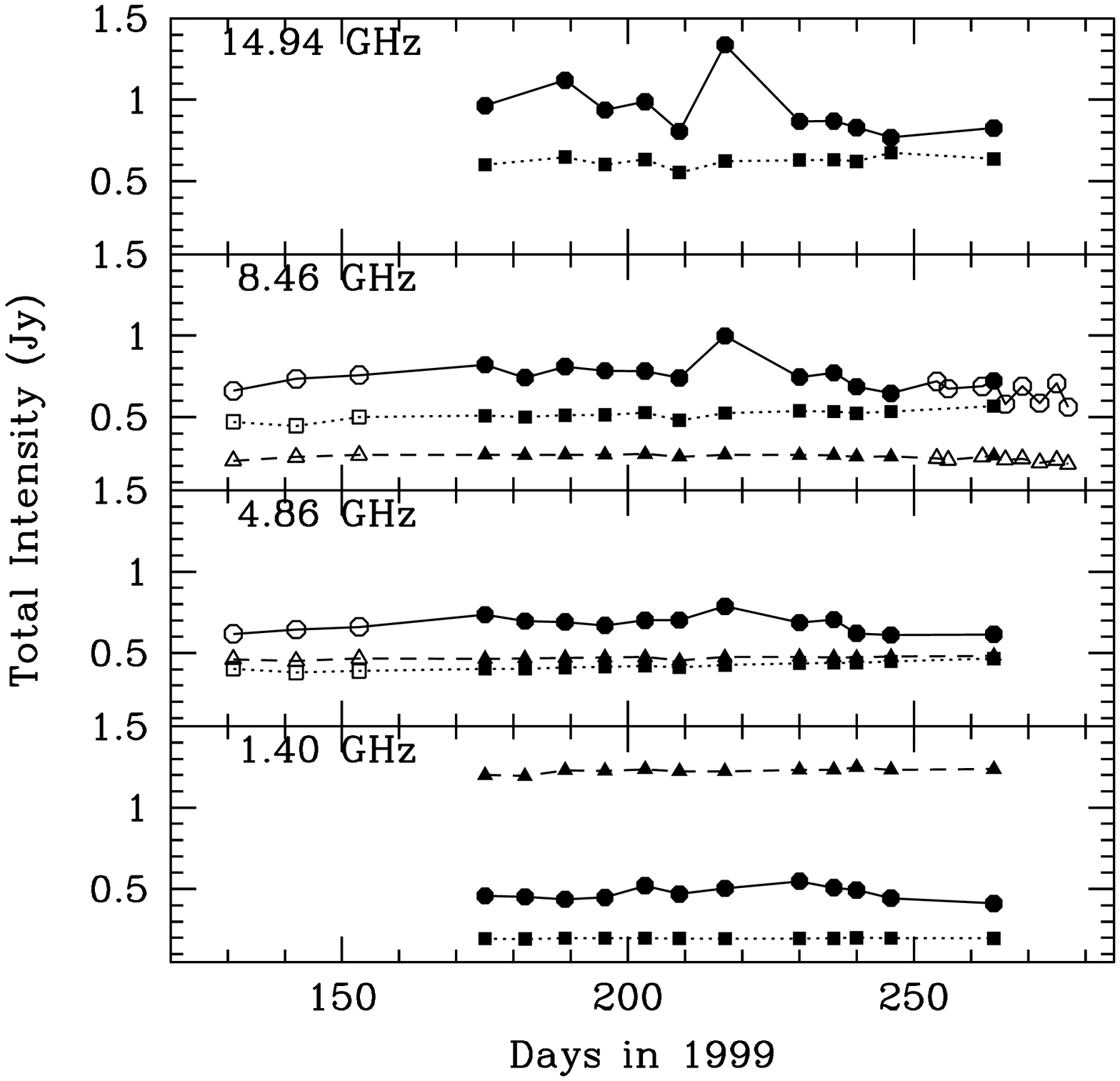}
\figcaption[f11.eps]{Total intensity for three
sources at 1.4, 4.8, 8.4 and 14.9 GHz from the 1999 VLA and ATCA
monitoring campaigns.
Symbols are the same as in Figure~10.}

\plotone{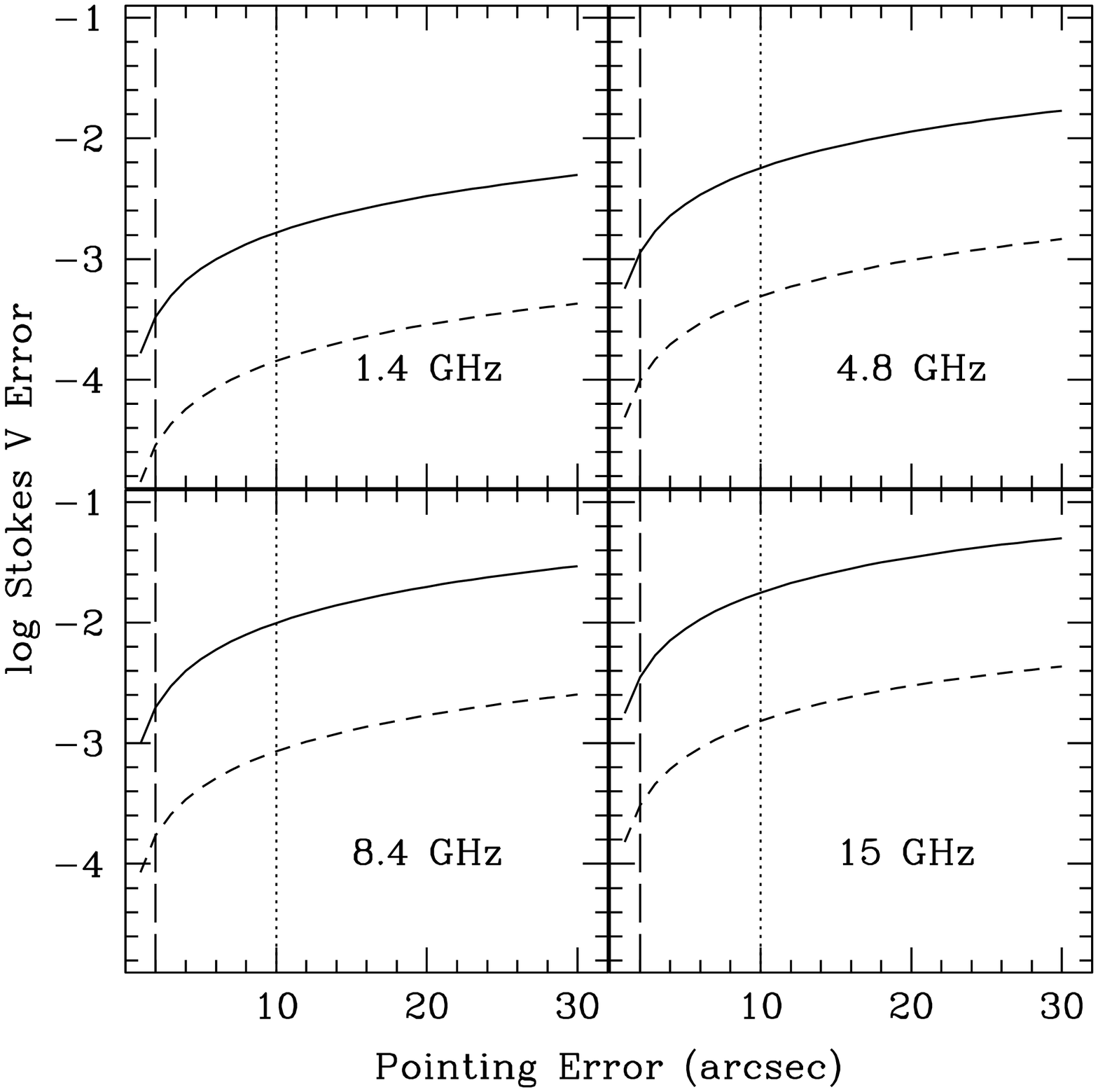}
\figcaption[f12.eps]{The fractional
error due to beam squint at four frequencies.
The solid curve is the error from a single antenna and the short-dashed
curve is the error from 27 antennas with 5 observations.  The dotted
vertical line is the expected rms pointing error without pointing
observations.  The long-dashed vertical line is the best-case rms pointing
error with pointing observations.}

\plotone{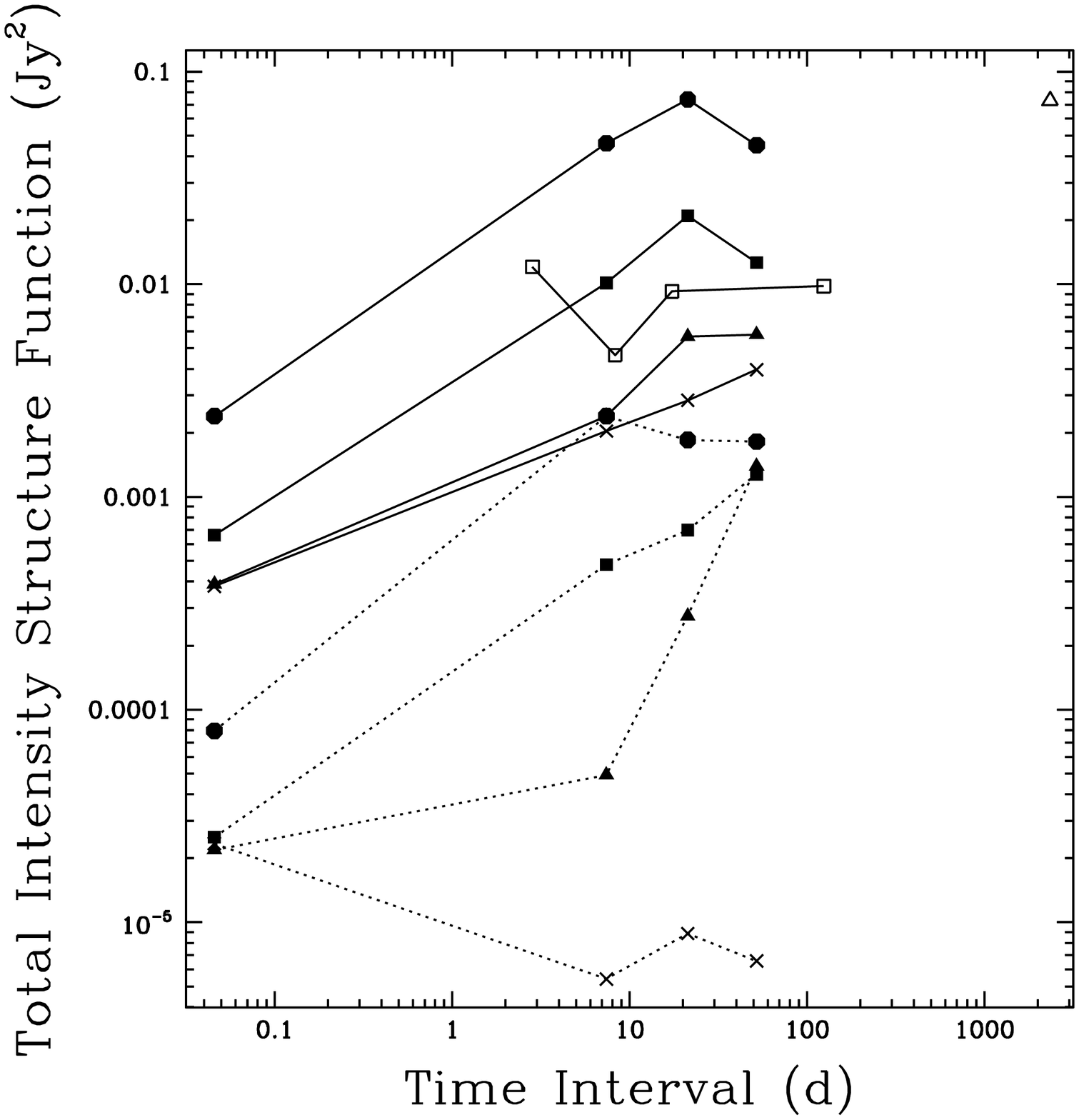}
\figcaption[f13.eps]{The structure function of the total intensity for Sgr A* 
(connected by solid lines)
and J1744-3116 (connected by dotted lines) from VLA
and ATCA measurements.  
The structure function derived solely from
VLA measurements in 1999 is indicated with crosses (1.4 GHz), filled
triangles (4.8 GHz), filled squares (8.4 GHz) and filled octagons
(15 GHz).  The structure function derived from VLA archive data
at 4.8 GHz is indicated with an open triangle.  The structure
function derived from ATCA measurements at 8.4 GHz is indicated
with open squares.  Errors for individual  structure function
measurements are on the order of a factor of two.\label{fig:isfun}}

\plotone{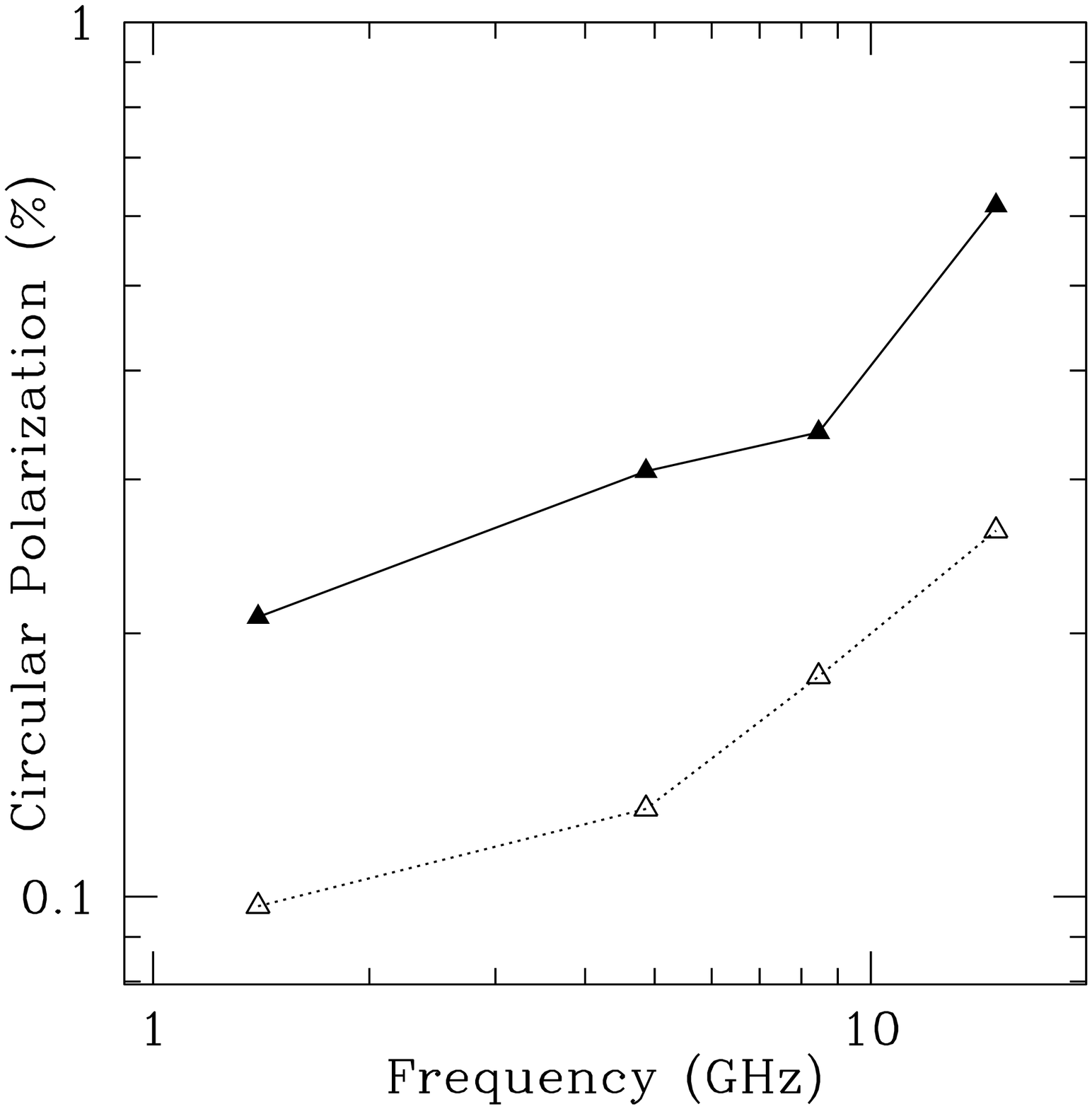}
\figcaption[f14.eps]{The mean spectrum (solid triangles) and degree of variability 
(open triangles) of fractional CP in Sgr A*
as a function of frequency.\label{fig:meanvar}}

\plotone{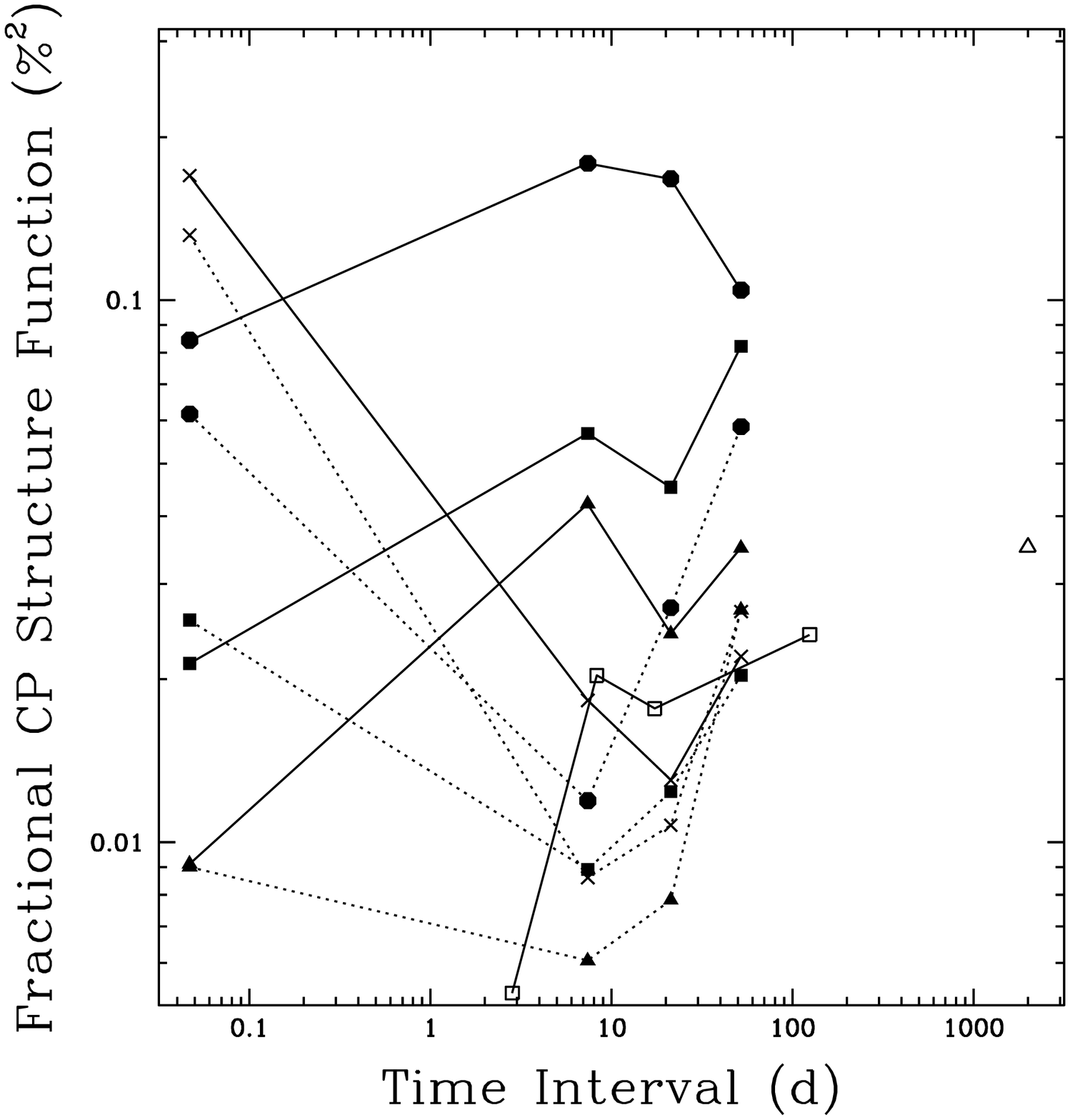}
\figcaption[f15.eps]{The structure function of fractional CP for 
Sgr A* and J1744-3116 from VLA
and ATCA measurements.
The units of fractional CP are in percent.  Thus,
a difference of 0.1\% corresponds to a structure function value of 0.01.
Symbols are the same as in Figure~\ref{fig:isfun}.
Errors for individual structure function measurements are on the 
order of a factor of two.\label{fig:sfun}}

\plotone{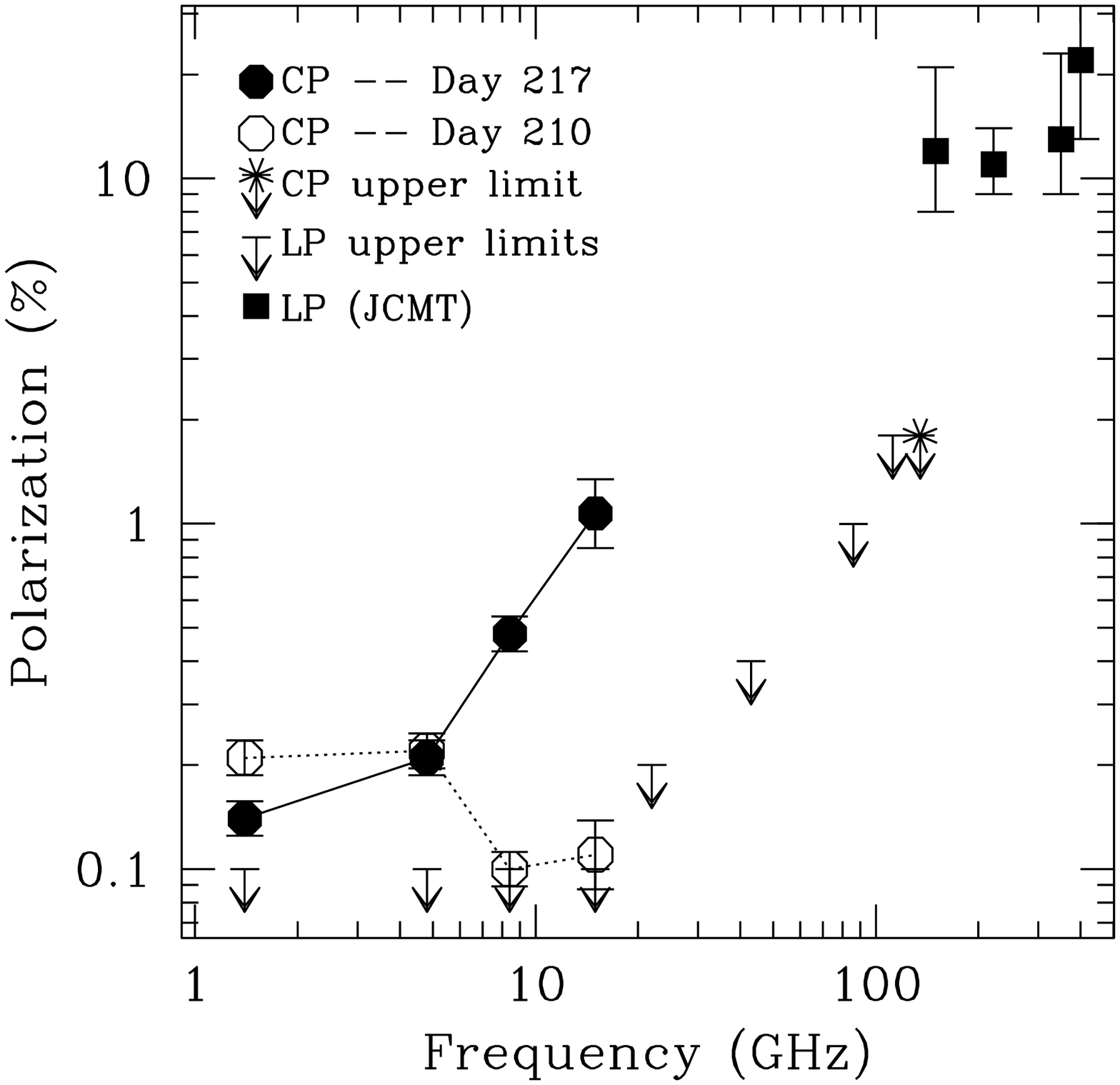}
\figcaption[f16.eps]{The spectrum of CP in Sgr A* from VLA measurements
on day 210 (open octagons) and day 217 (filled octagons).  We also
include an upper limit for CP from BIMA at 112 GHz 
\citep{2001ApJ...555L.103B}. 
LP upper
limits and reported JCMT detections (squares) are also plotted.  The sign of the CP has
been reversed for clarity.\label{fig:spectrum}}

\plotone{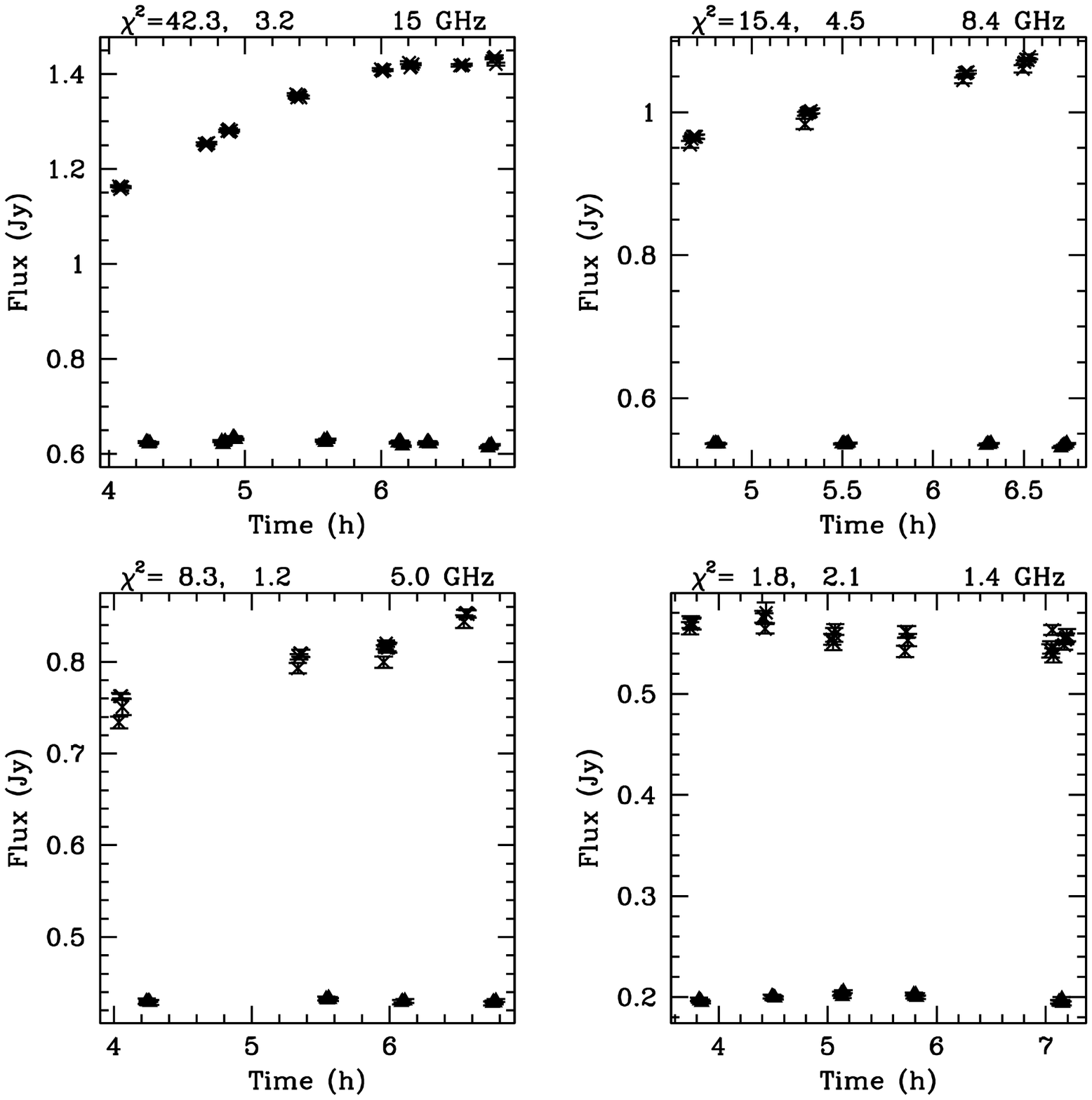}
\figcaption[f17.eps]{Short time scale variations in total intensity for Sgr A* (crosses)
and J1744-3116 (triangles) at 1.4, 4.8, 8.4 and 15 GHz from VLA observations
on day 217.  The reduced $\chi^2$ for the
hypothesis of constant flux density is listed in each box for Sgr A* and J1744-3116, respectively.
\label{fig:short0805x4}}

\plotone{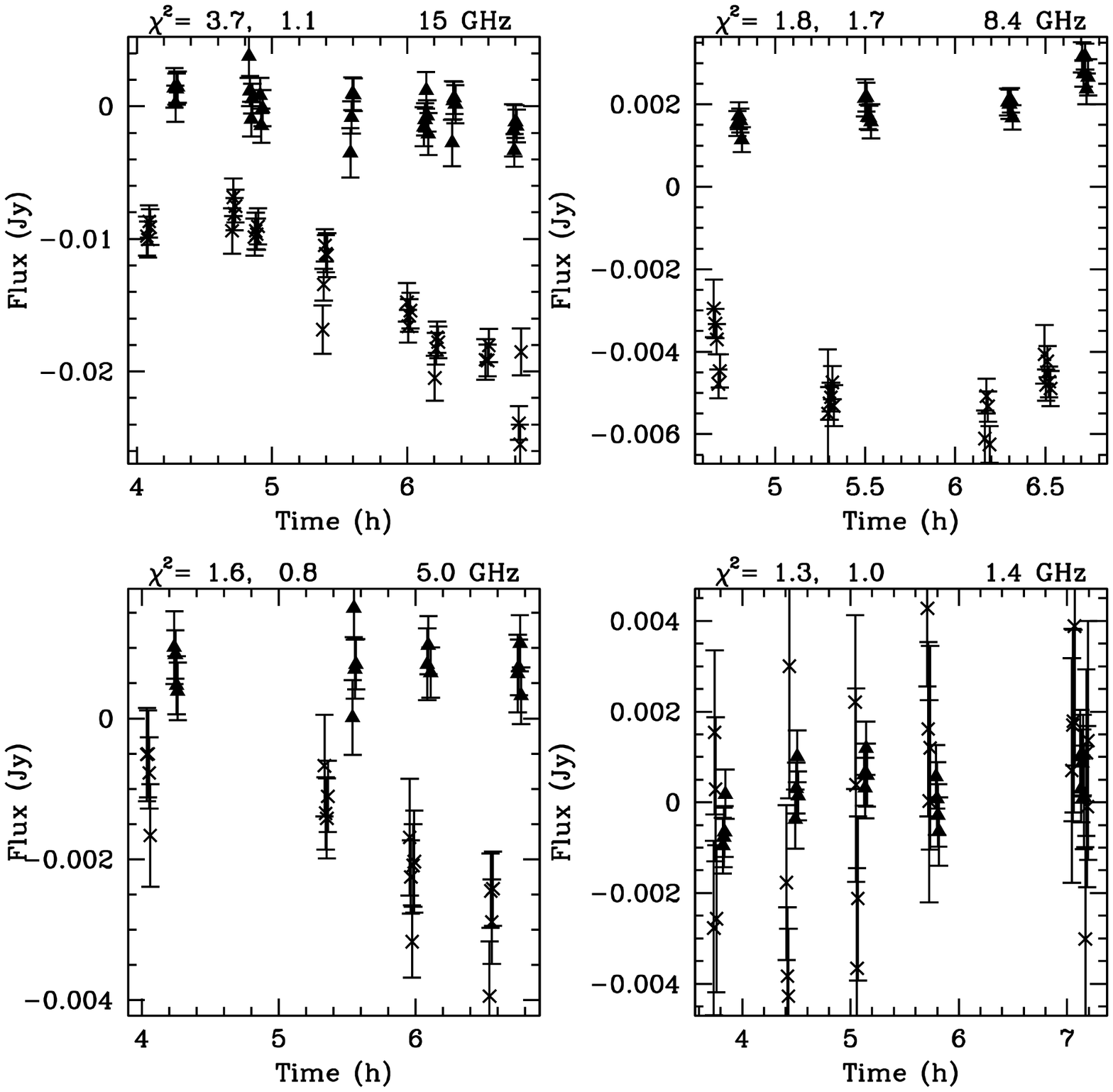}
\figcaption[f18.eps]{Short time scale variations in circular polarization for Sgr A* (crosses)
and J1744-3116 (triangles) at 1.4, 4.8, 8.4 and 15 GHz from VLA observations
on day 217.  The reduced $\chi^2$ for the
hypothesis of constant flux density is listed in each box for Sgr A* and J1744-3116, respectively.
\label{fig:short0805x4v}}

\begin{deluxetable}{llrrrr}
\tablecaption{Mean Total Intensity (Jy)\label{tab:meani}}
\tablehead{
\colhead{Source} & \colhead{Obs.} & \colhead{1.4 GHz} &
\colhead{4.8 GHz} & \colhead{8.4 GHz} & \colhead{15 GHz} }
\startdata
Sgr A* & VLA 1999 & $0.474 \pm 0.040$ & $0.685 \pm 0.051$ & $0.773 \pm 0.091$ & $0.937 \pm 0.166$ \\
       & ATCA 1999 & \dots           & $0.641 \pm 0.022$ & $0.669 \pm 0.065$ & \dots \\
       & VLA Arch &  \dots           & $0.868 \pm 0.191$ & $0.757 \pm 0.149$ & \dots \\
J1744-3116 & VLA 1999 & $0.196 \pm 0.002$ & $0.427 \pm 0.023$ & $0.520 \pm 0.030$ & $0.623 \pm 0.031$ \\
          & ATCA 1999 & \dots           & $0.391 \pm 0.012$ & $0.473 \pm 0.028$ & \dots \\
J1751-2523 & VLA 1999 & $1.226 \pm 0.015$ & $0.472 \pm 0.007$ & $0.264 \pm 0.006$ & \dots \\
	  & ATCA 1999 & \dots           & $0.459 \pm 0.008$ & $0.240 \pm 0.016$ & \dots \\
          & VLA Arch  & \dots           & $0.504 \pm 0.022$ & $0.276 \pm 0.003$ & \dots \\
J1820-2528 & ATCA 1999 & \dots           & $1.033 \pm 0.007$ & $1.025 \pm 0.062$ & \dots \\
J1733-1304 & ATCA 1999 & \dots           & $4.346 \pm 0.552$ & $3.798 \pm 0.257$ & \dots \\
J1743-038 & ATCA 1999 & \dots           & $4.842 \pm 0.528$ & $4.110 \pm 0.104$ & \dots \\
W56       & ATCA 1999 & \dots           & $0.092 \pm 0.004$ & $0.111 \pm 0.006$ & \dots \\
          & VLA Arch  & \dots           & $0.092 \pm 0.018$ & $0.132 \pm 0.022$ & \dots \\
W109      & ATCA 1999 & \dots           & $0.085 \pm 0.002$ & $0.070 \pm 0.003$ & \dots \\
          & VLA Arch  & \dots           & $0.090 \pm 0.011$ & $0.100 \pm 0.005$ & \dots \\
GC 441    & ATCA 1999 & \dots           & $0.042 \pm 0.002$ & $0.023 \pm 0.002$ & \dots \\
          & VLA Arch  & \dots           & $0.044 \pm 0.002$ & $0.026 \pm 0.001$ & \dots \\
\enddata
\end{deluxetable}

\begin{deluxetable}{llrrrr}
\tablecaption{Mean Fractional CP (\%)\label{tab:meancp}}
\tablehead{
\colhead{Source} & \colhead{Obs.} & \colhead{1.4 GHz} &
\colhead{4.8 GHz} & \colhead{8.4 GHz} & \colhead{15 GHz} }
\startdata
Sgr A* & VLA 1999 & $-0.21 \pm 0.10$ & $-0.31 \pm 0.13$ & $-0.34 \pm 0.18$ & $-0.62 \pm 0.26$ \\
       & ATCA 1999 & \dots           & $-0.37 \pm 0.08$ & $-0.27 \pm 0.10$ & \dots \\
       & VLA Arch &  \dots           & $-0.31 \pm 0.13$ & $-0.36 \pm 0.10$ & \dots \\
J1744-3116 & VLA 1999 & $-0.01 \pm 0.09$ & $-0.08 \pm 0.09$ & $0.24 \pm 0.09$ & $-0.12 \pm 0.14$ \\
          & ATCA 1999 & \dots           & $0.01 \pm 0.01$ & $0.04 \pm 0.05$ & \dots \\
J1751-2523 & VLA 1999 & $-0.08 \pm 0.08$ & $0.06 \pm 0.07$ & $0.04 \pm 0.09$ & \dots \\
	  & ATCA 1999 & \dots           & $0.00 \pm 0.02$ & $0.05 \pm 0.06$ & \dots \\
J1820-2528 & ATCA 1999 & \dots           & $-0.08 \pm 0.01$ & $0.08 \pm 0.04$ & \dots \\
J1733-1304 & ATCA 1999 & \dots           & $-0.14 \pm 0.09$ & $-0.10 \pm 0.06$ & \dots \\
J1743-0350 & ATCA 1999 & \dots           & $-0.10 \pm 0.02$ & $-0.01 \pm 0.05$ & \dots \\
W56       & ATCA 1999 & \dots           & $0.13 \pm 0.05$ & $0.22 \pm 0.11$ & \dots \\
W109      & ATCA 1999 & \dots           & $0.00 \pm 0.06$ & $-0.07 \pm 0.08$ & \dots \\
GC 441    & ATCA 1999 & \dots           & $0.01 \pm 0.08$ & $0.13 \pm 0.41$ & \dots \\
\enddata
\end{deluxetable}

\begin{deluxetable}{crrrcc}
\tablecaption{Estimated Errors in Fractional CP (\%)\label{tab:vnoise}}
\tablehead{
\colhead{$\nu$} & \colhead{Thermal} & \colhead{Beam Sq.} &
\colhead{Gain} & \colhead{D-terms} & \colhead{Total} \\
\colhead{(GHz)} &                   &                    &
               &                            & \colhead{($N=5$)} }
\startdata
1.4 & $0.02/\sqrt{N}$ &$0.04/\sqrt{N}$ &$0.06/\sqrt{N}$ & 0.03 & 0.05 \\
4.8 & $0.02/\sqrt{N}$ &$0.04/\sqrt{N}$ &$0.06/\sqrt{N}$ & 0.03 & 0.05 \\
8.4 & $0.02/\sqrt{N}$ &$0.04/\sqrt{N}$ &$0.06/\sqrt{N}$ & 0.03 & 0.05 \\
15  & $0.05/\sqrt{N}$ &$0.07/\sqrt{N}$ &$0.06/\sqrt{N}$ & 0.03 & 0.06 \\
\enddata
\end{deluxetable}

\begin{deluxetable}{crrr}
\tablecaption{Point-by-Point Variations in Measured VLA Fractional CP(\%)\label{tab:sigcp}}
\tablehead{
\colhead{$\nu$} & \colhead{Sgr A*} & \colhead{J1751-2523} &
\colhead{J1744-3116} \\ 
\colhead{(GHz)} &                   &                    & }
\startdata
1.4 & 0.15 & 0.07  & 0.10 \\
4.8 & 0.19 & 0.06  & 0.07 \\
8.4 & 0.22 & 0.03  & 0.10 \\
15  & 0.42 & \dots & 0.14 \\
\enddata
\end{deluxetable}

\end{document}